\documentclass[10pt,letterpaper]{article}
\usepackage[top=0.85in,left=2.75in,footskip=0.75in]{geometry}

\usepackage{amsmath,amssymb}

\usepackage{changepage}

\usepackage[utf8x]{inputenc}

\usepackage{array}
\usepackage{tabularx}
\usepackage{enumitem}
\usepackage{multirow}

\usepackage{times}
\usepackage{anyfontsize}

\usepackage{graphicx}
\graphicspath{{./figures/}}
\usepackage{caption}
\usepackage{float}
\usepackage{caption}
\usepackage{float}
\usepackage{tikz}
\usetikzlibrary{arrows}
\usetikzlibrary{calc}
\usepackage{subfig}
\usepackage{pgfplots}
\usetikzlibrary{shapes}
\usepackage{rotating}

\usepackage{multicol}
\usepackage{url}
\usepackage{color}
\usepackage{csquotes}
\usepackage{booktabs}
\usepackage[toc,page]{appendix}

\usepackage{cite} 
\usepackage[final]{hyperref} 
\hypersetup{
	colorlinks=true,       
	linkcolor=blue,        
	citecolor=blue,        
	filecolor=magenta,     
	urlcolor=blue         
}

\usepackage{microtype}
\DisableLigatures[f]{encoding = *, family = * }

\usepackage{xcolor}

\usepackage{array}
\usepackage{tabularx}
\usepackage{enumitem}
\usepackage{multirow}

\newcolumntype{+}{!{\vrule width 2pt}}

\newlength\savedwidth

\raggedright
\usepackage[aboveskip=1pt,labelfont=bf,labelsep=period,justification=raggedright,singlelinecheck=off]{caption}

\makeatletter
\renewcommand{\@biblabel}[1]{\quad#1.}
\makeatother

\usepackage{lastpage,fancyhdr,graphicx}
\usepackage{epstopdf}
\fancyheadoffset[L]{2.25in}
\fancyfootoffset[L]{2.25in}
\usepackage{ulem,cancel}

\newcommand{\ie}{{\it i.e.,~}}
\newcommand{\eg}{{\it e.g.,~}}
\newcommand{\AS}[1]{\textcolor{black}{#1}}

\begin{document}

\vspace*{0.2in}

\begin{flushleft}
{\Large
\textbf\newline{Optimal Regulation of Blood Glucose Level in Type I Diabetes using Insulin and Glucagon} 
}
\newline
\\
Afroza Shirin\textsuperscript{*},
Fabio Della Rossa\textsuperscript{},
Isaac S. Klickstein\textsuperscript{},
John J. Russell\textsuperscript{}, 
Francesco Sorrentino\textsuperscript{}
\\
\bigskip
Mechanical Engineering Department, University of New Mexico, Albuquerque, NM 87131
\bigskip

%
%





* ashirin@unm.edu

\end{flushleft}

\section*{Abstract}
\AS{The   Glucose-Insulin-Glucagon nonlinear model \cite{dalla2007meal,dalla2007gim,man2014uva,visentin2018uva}  accurately describes how the body responds to exogenously supplied insulin and glucagon in patients affected by Type I diabetes.}
Based on this model, we design infusion rates of either insulin (monotherapy) or insulin and glucagon (dual therapy) that can optimally maintain the blood glucose level within desired limits after consumption of a meal and prevent the onset of both hypoglycemia and hyperglycemia.
This problem is formulated as a nonlinear optimal control problem, which we solve using the numerical optimal control package $\mathcal{PSOPT}$.
Interestingly, in the case of monotherapy, we find the optimal solution is close to the standard method of insulin based glucose regulation, which is to assume a variable amount of insulin half an hour before each meal.
We also find that the optimal dual therapy (that uses both insulin and glucagon) is better able to regulate glucose as compared to using insulin alone.
\AS{We also propose an \textit{ad-hoc} rule for both the dosage and the time of delivery of insulin and glucagon.}



\section{Introduction}
Insulin and glucagon are pancreatic hormones that help regulate the levels of glucose in the blood.
Insulin is produced by  the {\it beta-cells} in the pancreas and carries glucose from the bloodstream to the cells throughout the body.
Glucagon releases glucose from the liver into the bloodstream in order to prevent hypoglycemia. 
In people affected by diabetes 
insulin is either absent (type I diabetes) or not produced in the proper amount (type II diabetes).
In type I diabetes 
the body’s immune system attacks and destroys the beta cells.
As a result, insulin is not produced and glucose accumulates in the blood which may cause serious harm to several organs.  
Type II diabetes is a metabolic disorder in which the beta cells are unable to properly regulate the blood glucose within limits.
Common therapies for diabetes involve the administration of exogenous insulin.
Currently glucagon is not typically included in therapies because it does not preserve its chemical properties at room temperature and also because diabetic patients are still able to produce it.

The control of glucose levels in diabetic patients is an active field of research
\cite{steil2006feasibility,magni2008evaluating,magni2009model,palerm2008run,bergenstal2013threshold,van2014feasibility,nimri2014md,capel2014artificial,mauseth2015stress,reddy2014feasibility,parker1999model,parker2001intravenous,gillis2007glucose,lynch2001estimation,bruttomesso2009closed}. 
The approval by the FDA of a simulator which replaces {\it in-vivo} with {\it in-silico} therapy testing has greatly benefited this area of research.
This simulator implements a mathematical model, first proposed in \cite{dalla2007meal} and updated in \cite{dalla2007gim,man2014uva,visentin2018uva}, and provides an alternative to often slow, dangerous and expensive human testing.

Typically, insulin is administered manually approximately half an hour before each meal where the amount is determined from the current glucose level (measured through a blood sugar test), the expected glucose intake, and the patient's sensitivity to insulin.
In what follows we will refer to this as the \textit{standard therapy}.
In 1992 the first insulin pumps were introduced to the market.
They delivered both a consistent basal amount of insulin and an insulin bolus determined by the patients based on their glucose level.
It was only in 2016 that the first autonomous system for glycemic control was approved by the FDA.
The system consists of an insulin pump, a sensor that measures the blood glucose level continuously in time, and control software that is able to regulate the insulin level in the blood without needing any input from the patient.

Many control techniques have been proposed and tested to regulate blood glucose levels using insulin pumps including PID (proportional–integral–derivative) control \cite{steil2006feasibility,magni2008evaluating,palerm2008run,bergenstal2013threshold,van2014feasibility,marchetti2008improved}, fuzzy logic control 
\cite{nimri2014md,capel2014artificial,mauseth2015stress} and bio-inspired techniques \cite{reddy2014feasibility} which do not rely on a mathematical model. 
In \cite{fisher1991semiclosed} closed loop control has been used on a so called ``minimal model" \cite{bergman1979quantitative,bergman1981physiologic,bergman1985assessment}.
In \cite{parker1999model,parker2001intravenous,gillis2007glucose,lynch2001estimation,thabit2015home} a linear model predictive control (MPC) has been used in a model with fixed structure but for which parameters are constantly updated to adapt to the patient's response. \AS{In \cite{zavitsanou2015silico}  linear MPC has been used \textit{in silico}.}
In \cite{hovorka2004nonlinear} MPC has been applied to a system linearized around the operating points of a physically derived nonlinear model and in \cite{bequette2012challenges} multiple model probabilistic predictive control has been used.
In \cite{copp2018simultaneous} MPC has been applied together with a moving horizon estimation technique to a linear model.  
Most of the models used when designing the above controllers are simplified versions of the FDA approved model and all the control techniques considered only use insulin (but not glucagon) as control input.

Because insulin delivered exogenously is not subject to normal physiological feedback regulation, hypoglycemia is common in patients with Type 1 diabetes who undergo treatment \cite{mccrimmon2010hypoglycemia}.
\AS{For these patients it has been proposed that exogenous insulin can be used to lower their blood glucose level and exogenous glucagon can be used to prevent hypoglycemia \cite{castle2010novel,batora2015contribution}.
Currently, a commercial pump that delivers both insulin and glucagon is not available, and the development of a two-drug artificial pancreas is still the subject of clinical research \cite{el2007adaptive,el2010bihormonal,russell2012blood,el2014autonomous,russell2014outpatient,russell2016day,el2017home,herrero2017coordinated,boiroux2018adaptive}.}
An outstanding research question, which we address in this paper, is the determination of the temporal dosages of both insulin and glucagon, in the case of the dual therapy. 

Following the study in \cite{shirin2018prediction} which optimized multi-drug therapies for autophagy regulation, here we seek to determine an optimal strategy for delivery of both insulin and glucagon.
We consider the combined effects of insulin and glucagon in regulating blood glucose levels in patients with Type 1 diabetes, using the  model in \cite{man2014uva} and nonlinear optimal control theory.
Additionally, the objective function that we seek to minimize is the Blood Glucose Index which is a well known tool to measure the risk for a patient to enter either hyperglycemia or hypoglycemia.
To design the optimal control problem, we use the balance control technique of ref.\ \cite{shirin2017optimal}, which introduces a trade-off between the error allowed with respect to a state based cost (Blood Glucose Index) and the control effort.
Our goal is to evaluate the performance limits of a control algorithm in the blood glucose problem, and to discuss the advantages of the dual drug therapy compared to the  single drug therapy.
Note that even though we do not attempt to design a closed-loop control strategy that works without the patient's intervention, the solution we propose can be adapted for that purpose.

From solving the optimal control problem for a family of objective functions derived from the balance control paradigm, we observe the emergence of a pattern,  from which we propose a simple rule for the delivery of insulin and glucagon similar to the standard therapy, but for the case that both insulin and glucagon are used.
While this therapy is suboptimal, we see that it still performs better than the optimal solution with insulin alone.

\AS{Finally, we test the robustness of the optimal solution.
While optimal control does not guarantee robustness of the optimal solution with respect to model uncertainty or parameter mismatches, we see that our proposed solution still performs well in the presence of model parameter perturbations and variations affecting the time and glucose intake of the meal.}

\section{Materials and methods}

\subsection{Model and Parameters}
We consider the model in \cite{man2014uva,visentin2018uva} which is a system of nonlinear ordinary differential equations (ODEs).
The equations are given in Eqs.\ \eqref{eq:ode1}-\eqref{eq:ode9} in supplementary information section \ref{sec:model}. 
%
We write the  ODEs in Eqs.\ \eqref{eq:ode1}-\eqref{eq:ode9} in the form 
\begin{equation}\label{eq:ode}
  \begin{array}{rcl}
    \dot{\textbf{x}}(t) &=& \textbf{f}(\textbf{x}(t),\textbf{u}(t),{D(t)}, \Theta)\\
    G(t)&=&x_1/V_G
  \end{array}
\end{equation}
where the state vector is $\textbf{x} =[x_1(t), x_2(t),..., x_{17}(t)]^T$ and $t$ is the physical time (in min).
In Table \ref{tab:variables} we tabulate all of the variables $x_i$ and their names.
The control input vector is $\textbf{u}(t) = [u_I(t), u_G(t)]^T$, where $u_I(t) \ge 0$ is the exogenous insulin infusion rate (in insulin Unit/min) and $u_G(t) \ge 0$ is the exogenous glucagon infusion rate (mg/min).
Both $u_I(t)$ and $u_G(t)$ are the external inputs to the system in Eq.\ \eqref{eq:ode}.
The scalar quantity $D(t)$ represents the exogenous glucose input, that is, the glucose intake with a meal.
The output of the system is the quantity $G(t)$, which measures the density of glucose in the blood, obtained as the ratio between the plasma glucose and the distribution volume of glucose $V_G$.
%

When $u_I(t) = 0$, $u_G(t) = 0$ and $D(t) = 0$, the model reaches (for physically meaningful parameters) a steady state, also known as the {\it basal condition} of a patient.
The basal condition depends upon the parameters of the models $\Theta$.
We denote by  $\Theta_{G_b}$ a set of parameters for which the basal glucose level {$G$}  is equal to $G_b$. 
The basal levels for the other states are found according to Eqs.\ \eqref{eq:basal}. 


\begin{table}[htbp]
	\begin{adjustwidth}{-2.25in}{0in}
		\centering
		\caption{Variables and their physical meaning}
		{\setlength\doublerulesep{2pt}   
			\begin{tabular}{llll}
				\toprule[1pt] \midrule
				Variables  & Names & Representing & Units \\
				\midrule \midrule 
				$x_1$  & $G_p$ &  Mass of  glucose in plasma & 	 mg/kg	\\
				$x_2$  & $G_t$ &  Mass of  glucose in tissue &	mg/kg	\\
				$x_3$  & $I_l$ & Mass of  insulin in liver & pmol/kg	 	\\
				$x_4$  & $I_p$ & Mass of  insulin in plasma & pmol/kg	 	\\
				$x_5$  & $I'$ & Mass of  delayed in compartment 1 & pmol/L		\\
				$x_6$  & $X^L$ & Amount of delayed	insulin action on $EGP$ (Endogenous glucose production) & pmol/L	\\
				$x_7$  & $Q_{sto1}$ & Amount of solid glucose in stomach & mg \\
				$x_8$  & $Q_{sto2}$ & Amount of liquid glucose in stomach &  mg\\
				$x_9$  & $Q_{gut}$ & Amount of glucose in intestine & mg	 	\\
				$x_{10}$  & $X$ & Amount of interstitial fluid	&  pmol/L	\\
				$x_{11}$  & $SR^s_H$ & Amount of static glucagon &	ng/L/min	\\
				$x_{12}$  & $H$ & Amount plasma glucagon & ng/L\\
				$x_{13}$  & $X^H$ & Amount of delayed glucagon	action on $EGP$	& ng/L\\
				$x_{14}$  & $I_{sc1}$ & Amount of nonmonomeric insulin in the subcutaneous space & pmol/kg \\
				$x_{15}$  & $I_{sc2}$ & Amount of monomeric insulin & pmol/kg \\
				$x_{16}$  & $H_{sc1}$ & Amount of glucagon in the subcutaneous space 1 & ng/L	\\
				$x_{17}$  & $H_{sc2}$ & Amount of glucagon in the subcutaneous space 2	& ng/L\\				
				\midrule \bottomrule[1pt]
				
			\end{tabular}
		}
		\begin{flushleft}State variables and their physical meaning.
		\end{flushleft}
		\label{tab:variables}
	\end{adjustwidth}
\end{table}

\subsection{Problem Formulation}

We formulate a nonlinear optimal control problem with two control goals.
The first goal is to regulate the glucose at levels corresponding to low clinical risk of either hyperglycemia or hypoglycemia during a time period over which a meal is consumed.
We assume that a meal is ingested at time $t=\tau_D$, which we assume to be modeled as a Dirac delta function $D(t)=D \delta(t-\tau_D)$. 
To evaluate the clinical risk of a particular glycemic value, Kovatchev et al. \cite{kovatchev1997symmetrization,kovatchev2005quantifying} proposed the {\it Blood Glucose Index} (BGI), defined as
\[
BGI\left(G(t)\right) = 10\left(1.509\left((\ln{G(t)})^{1.084}-5.3811\right)\right)^2,
\]
where a small BGI value corresponds to low risk of either hyperglycemia or hypoglycemia.
This metric also takes into account the fact that (i) the target blood glucose range as defined by the Diabetes Control and Complications Trial \cite{diabetes1993effect} (between 70 and 180 mg/dL) is not symmetric about the center of the range and (ii) hypoglycemia occurs at glucose levels closer to the basal level than hyperglycemia. 
%
The second goal is to limit the overall usage of insulin and/or glucagon over the period $[t_0,t_f]$.

We formulate the optimization problem according to these two goals,
\begin{equation}\label{eq:obj}
  \begin{aligned}
    \min_{\textbf{u}(t)} \quad  J = &   \int_{t_0}^{t_f}  \left[ \alpha_p BGI\left(G(t)\right) + \alpha_I u_I^p(t) + \alpha_G u_G^p(t)\right] dt,
  \end{aligned}
\end{equation}
subject to the following constraints,
\begin{subequations}\label{eq:const}
	\begin{align}
	&& &\begin{array}{l}
	\dot{\textbf{x}}(t) = \textbf{f} (\textbf{x}(t),\textbf{u}(t),{D \delta(t-\tau_D),}\Theta_{G_b}), \quad  \textbf{u}(t) =  [u_I(t)\quad u_G(t)]^T	\label{eq:ode_fx}
	\end{array}\\[0.05in] 
	&& &\begin{array}{l}
	G^L < G(t) < G^U  \label{eq:glucoselevel}
	\end{array}\\[0.05in] 
	&& &\begin{array}{l}
	u_I^L \le u_I(t) \le u_I^U  \label{eq:insulinlevel}
	\end{array}\\[0.05in] 
	&& &\begin{array}{l}
	0 \le u_G(t) \le u_G^U \label{eq:gluccagonlevel}
	\end{array}\\[0.05in] 
	&& &\begin{array}{l}
	0 \le \int_{t_0}^{t_f} u_I(t) dt \le \phi_I^U  \label{eq:phiI}
	\end{array}\\[0.05in]
	&& &\begin{array}{l}
	0 \le \int_{t_0}^{t_f} u_G(t) dt \le \phi_G^U  \label{eq:phiG}
	\end{array}\\[0.05in]
	&& &\begin{array}{l}
	\textbf{x}(t_0) = \bar{\textbf{x}} \label{eq:initialcondition}
	\end{array}
	\end{align}
\end{subequations}
In Eqs.\ \eqref{eq:obj} and \eqref{eq:const}, the insulin infusion rate $u_I(t)$ and  the glucagon infusion rate $u_G(t)$ are the two control inputs.
The three coefficients $\alpha_p$, $\alpha_I$ and $\alpha_G$ in Eq. \eqref{eq:obj} are tunable factors through which we may vary the weight associated with each of the three terms in the cost function $J$. 
The first coefficient, $\alpha_p$ is dimensionless while the units of $\alpha_I$ and $\alpha_G$ are $(\text{U/min})^{-p}$ and $(\text{mg/min})^{-p}$, respectively. 
Note that by setting $u_G = 0$ in Eq.\ \eqref{eq:gluccagonlevel}, we have an optimal control problem in terms of insulin only.

The first term in the objective function \eqref{eq:obj} defines a regulation problem, \ie we try to maintain the glucose at low risk levels.
The second and third terms in the cost function are chosen to avoid using excess insulin or glucagon.
For $p=1$ in Eq. \eqref{eq:obj}, the second and third terms define a {\it `minimum fuel'} problem, thus we call the optimization problem ReMF (Regulation and Minimum Fuel).
In this case, we expect the optimal solution to consist of pulsatile inputs $u_I^*(t)$ and $u_G^*(t)$ \cite{kirk2012optimal,chachuat2007nonlinear}. 
For $p=2$, the second and third term inside the cost function define a {\it `minimum energy'} problem, thus we call the optimization problem ReME (Regulation and Minimum Energy).
In this case, we expect the optimal control inputs $u_I^*(t)$ and $u_G^*(t)$ to be continuous. 
The set of equations in  \eqref{eq:ode_fx} coincide with the ODEs in Eqs.\ \eqref{eq:ode1}-\eqref{eq:ode9}of the supplemental information.
In Eq.\ \eqref{eq:glucoselevel} $G^L$ and $G^U$ are the lower and upper bounds for $G(t)$,
they can be set in order to avoid undesired hypoglycemic or hyperglycemic states.
In Eqs.\ \eqref{eq:insulinlevel} and \eqref{eq:gluccagonlevel} $u_I^U$ and $u_G^U$ are upper bounds for the insulin and glucagon delivery rates, respectively.
These constraints are set by the maximum infusion rates allowed by the insulin pump.
In Eq.\ \eqref{eq:insulinlevel} $u_I^L \geq 0$ is the lower bound for $u_I(t)$, \ie a minimum insulin delivery rate that can be used to set a basal insulin infusion rate to counteract endogenous  glucose production \cite{jdrfwebsite}.
Finally, in Eqs.\ (\ref{eq:phiI}, \ref{eq:phiG}), $\phi_I^U$ and $\phi_G^U$ set limits to the total limits of insulin and glucagon that can be delivered over the time period $[t_0,t_f]$.
The initial condition $\bar{\textbf{x}}$ in Eq.\ \eqref{eq:initialcondition} defines the patient's condition before administration of the therapy.
In the Results section, we discuss how we choose the bounds on $G(t)$, $u_I(t)$, $u_G(t)$, $\phi_I$, $\phi_G$, the control time period $[t_0,t_f]$ and the initial condition $\bar{\textbf{x}}$. 

Our goal is to find an optimal solution which satisfies the constraints in Eq.\ \eqref{eq:const} and minimizes the objective function \eqref{eq:obj}.
Note that the BGI only depends upon $G(t)$:
we are making no attempt to control the states of the system, only its output.
In the literature, such an approach is often referred to as target control \cite{kirk2012optimal,klickstein2017energy}.

\subsection{Method: Pseudo-Spectral Optimal Control}
Optimal control theory combines aspects of dynamical systems, optimization, and the calculus of variations \cite{kirk2012optimal} to solve the problem of finding a control law for a given dynamical system such that the prescribed optimality criteria are achieved.
The equations \eqref{eq:obj} and \eqref{eq:const} together form a constrained optimal control problem, which can generally be written as,
\begin{equation}\label{eq:OCP}
  \begin{aligned}
    \min_{\textbf{u}(t)} && &J(\textbf{x}(t),\textbf{u}(t),t) = \int_{t_0}^{t_f} F\left( \textbf{x}(t), \textbf{u}(t), t\right) dt\\
    \text{s.t.} && &\dot{\textbf{x}}(t) = \textbf{f} ( \textbf{x}(t), \textbf{u}(t), t)\\
    && &\textbf{e}^L \leq \textbf{e}(\textbf{x}(t_0), \textbf{x}(t_f), t_0, t_f) \leq \textbf{e}^U\\
    && &\textbf{h}^L \leq \textbf{h}(\textbf{x}(t), \textbf{u}(t), t) \leq \textbf{h}^U\\
    && &t \in [t_0,t_f]
  \end{aligned}
\end{equation}
In general, there exists no analytic framework that is able to provide the optimal time traces of the controls $\textbf{u}^*(t)$ and the states $\textbf{x}^*(t)$ in \eqref{eq:OCP}, and so we must resort to numerical techniques.

Pseudo-Spectral Optimal Control (PSOC) is a computational method for solving optimal control problems. Here we present a brief overview of the theory of pseudo-spectral optimal control.
PSOC has become a popular tool in recent years \cite{rao2009survey, ross2012review} that has let scientists and engineers solve optimal control problems like \eqref{eq:OCP} reliably and efficiently in applications such as guiding autonomous vehicles and maneuvering the international space station \cite{ross2012review}.
PSOC is an approach by which an OCP can be discretized by approximating the integrals by quadratures and the time-varying states and control inputs with interpolating polynomials.
Here we summarize the main concept of the PSOC. 
We choose a set of $N$ discrete times $\{\tau_i\}$ $i = 0,1,\ldots,N$ where $\tau_0 = -1$ and $\tau_N = 1$ with a mapping between $t \in [t_0,t_f]$ and $\tau \in [-1,1]$.
The times $\{\tau_i\}$  are chosen as the roots of an $(N+1)$th order orthogonal polynomial such as Legendre polynomials or Chebyshev polynomials.
The choice of dicretization scheme is important to the convergence of the full discretized problem.
For instance, if we choose the roots of a Legendre polynomial as the discretization scheme, the associated quadrature weights can be found in the typical way for Gauss quadrature.
The time-varying states and control inputs are found by approximating them with Lagrange interpolating polynomials,
\begin{subequations}
    \begin{align}
    \hat{\textbf{x}}(\tau) & = \sum_{i=0}^N \hat{\textbf{x}}_i L_i(\tau)\\
    \hat{\textbf{u}}(\tau) & = \sum_{i=0}^N \hat{\textbf{u}}_i L_i(\tau),
    \end{align}
\end{subequations}
where $\hat{\textbf{x}}(\tau)$  and  $\hat{\textbf{u}}(\tau)$ are the approximations of $\textbf{x}(\tau)$ and $\textbf{u}(\tau)$, respectively, and $L_i(\tau)$ is the $i$th Lagrange interpolating polynomial.
The dynamical system is approximated by differentiating the approximation $\hat{\textbf{x}}(\tau) = \sum_{i=0}^N \hat{\textbf{x}}_i L_i(\tau)$ with respect to time.
\begin{equation}
\begin{aligned}
\frac{d \hat{\textbf{x}}}{d \tau} = \sum_{i=0}^N \hat{\textbf{x}}_i \frac{d L_i}{d\tau}
\end{aligned}
\end{equation}
Let $D_{k,i} = \frac{d}{d\tau} L_i(\tau_k)$ which allows one to rewrite the original dynamical system constraints in \eqref{eq:OCP} as the following set of algebraic constraints.
\begin{equation}
\sum_{i=0}^N D_{k,i} \textbf{x}_i - \frac{t_f-t_0}{2} \textbf{f}(\hat{\textbf{x}}_i,\hat{\textbf{u}}_i,\tau_i) = \boldsymbol{0},\ k = 1,\ldots,N
\end{equation}
%
%
%
The integral in the cost function is approximated as,
	\begin{equation}
	J = \int_{t_0}^{t_f} F(\textbf{x},\textbf{u},t) \approx \hat{J} = \frac{t_f-t_0}{2} \sum_{k=1}^N F(\hat{\textbf{x}}_k, \hat{\textbf{u}}_k, \tau_k)
	\end{equation}
The original time-varying states, control inputs, the dynamical equations constrained and the cost function are now discretized approximation of the continuous NLP problem.
Thus the discretized approximation of the original OCP is compiled into the following nonlinear programming (NLP) problem.
\begin{equation}\label{eq:dOCP}
\begin{aligned}
\min_{\substack{\textbf{u}_i\\ i=0,\ldots,N}} && &\hat{J} = \frac{t_f-t_0}{2} \sum_{i=0}^N w_i f(\hat{\textbf{x}}_i,\hat{\textbf{u}}_i,\tau_i)\\
\text{s.t.} && &\sum_{i=0}^N D_{k,i} \hat{\textbf{x}}_i - \frac{t_f-t_0}{2} \textbf{f}(\hat{\textbf{x}}_k,\hat{\textbf{u}}_k,\tau_k) = \boldsymbol{0},\ k = 0,\ldots,N\\
&& &\textbf{e}^L \leq \textbf{e}(\hat{\textbf{x}}_0,\hat{\textbf{x}}_N,\tau_0,\tau_N) \leq \textbf{e}^U\\
&& &\textbf{h}^L \leq \textbf{h}(\hat{\textbf{x}}_k,\hat{\textbf{u}}_k, \tau_k) \leq \textbf{h}^U,\ k = 0,\ldots,N\\
&& &t_i = \frac{t_f-t_0}{2}\tau_i + \frac{t_f+t_0}{2}
\end{aligned}
\end{equation}
We have used $\mathcal{PSOPT}$ \cite{becerra2010solving}, an open-source PSOC library, to perform the above PSOC discretization procedure.
The NLP in \eqref{eq:dOCP} can be solved with a number of different techniques, but here we use an interior point algorithm \cite{nocedal2006numerical} as implemented in the open-source software Ipopt \cite{wachter2006implementation}.

\subsubsection{Continuous Approximation of Non-differential Function in ODEs}
\AS{The optimization algorithms implemented in $\mathcal{PSOPT}$ require the derivatives of the function $\textbf{f}(\textbf{x}(t),\textbf{u}(t), \Theta_{G_b})$ exists.
As there are terms that contain discontinuities in Eqs.\ \eqref{eq:ode1}-\eqref{eq:ode9}, we replace them with smooth approximations which are described in  section \ref{sec:approx} of the SI.}

\section{Results}

\begin{figure}[t!]
\begin{adjustwidth}{- 0.5 in}{0in} 
	\includegraphics[width= 1.1\textwidth]{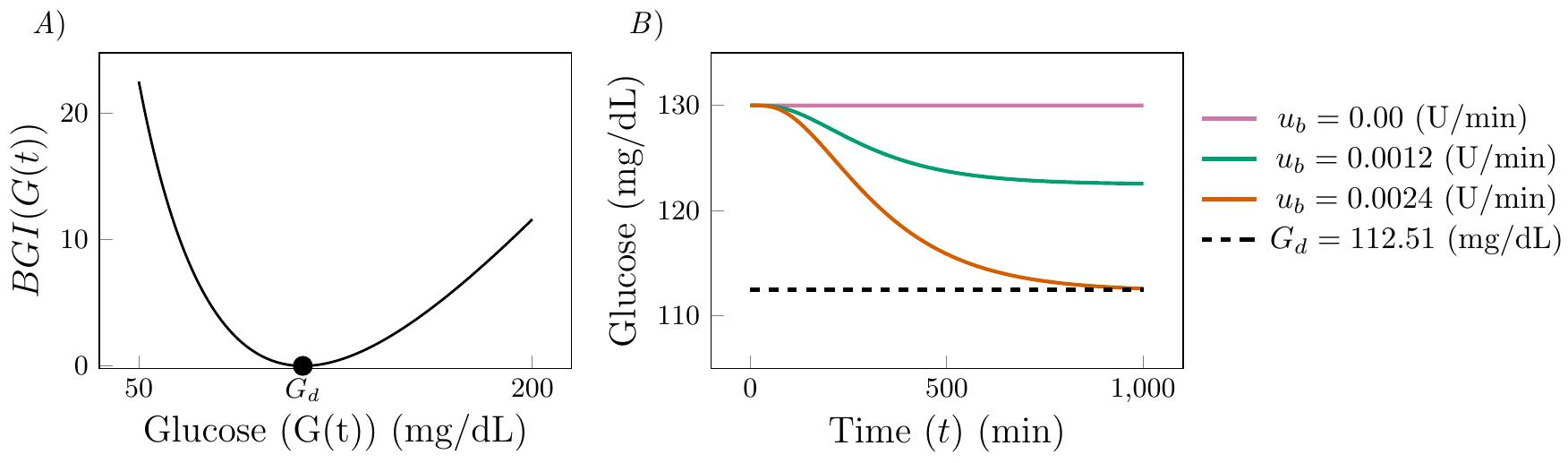}
	\caption{\textit{A}) The Blood Glucose Index ($BGI(G(t))$) as a function of the blood glucose $G(t)$. The function is minimized at $G(t) = G_d = 112.51$ (mg/dL). \textit{B}) The response of glucose ($G(t)$) to different time-constant basal insulin infusion rates in the absence of a meal.
	We see that as $u_b$ increases, the glucose is further down regulated.}
	\label{fig:constinput_1}
	\end{adjustwidth}
\end{figure}

We now describe in more detail the optimal control problem in Eqs. \ref{eq:obj} and \ref{eq:const} by setting the constraint and parameter values.
In Fig.\ \ref{fig:constinput_1}{\it A} we plot the function $BGI(G)$ versus the glucose $G$.
The minimum $BGI\left(G\right)$ occurs at $G = G_{d} = 112.51$ mg/dL,  which corresponds to a clinical target set for the glucose level \cite{diabetes1993effect}.
\AS{Based on the data in \cite{russell2015insulin}, the average fasting plasma glucose level of patients with type I diabetes is $G_b = 130$ (mg/dL). Thus, we set the the basal glucose level $G_b = 130$ (mg/dL).
The parameters $\Theta_{G_b}$ are set so that the steady state glucose is $130$ (mg/dL) in the absence of a meal and of exogenously supplied insulin, \ie we compute $\Theta_{130}$.}

We set the upper and lower bounds for the glucose level, $G^L$ and $G^U$ in Eq. \eqref{eq:glucoselevel}, to satisfy the target blood glucose range, $90\le G(t) \le 180$ \cite{diabetes1993effect}. 
The control time period is $[t_0,t_f] = [0,300]$ minutes, and we assume that a meal with 70 grams of glucose is consumed at time $t = 60$ min (\ie $D(t)=70\delta(t-60)$). 

We consider a situation in which the patient's glucose level is partially controlled by providing a constant but low insulin infusion rate $u_b>0$ (which is common for patients who use an insulin pump) \cite{jdrfwebsite} and serves to compensate for the endogenous glucose production.
In figure \ref{fig:constinput_1}{\it B} we show glucose response $G(t)$ for different values of constant $u_b$ in the absence of a meal.
\AS{We observe that for $u_b = 0.0024$ (U/min), $G(t)$ converges to the desired glucose level $G_d$.}
We thus set the lower bound of $u_I(t)$ in Eq.\ \eqref{eq:insulinlevel}, $u_I^L = u_b$, while its upper bound is set to $u_I^U = 15 $ (in U/min), the maximum insulin flow allowed in commercial pumps \cite{userguide}.
In the absence of commercially available glucagon pumps, we will assume that a pump mechanically similar to an insulin pump is used to deliver glucagon.
Since the maximum flow rate for an insulin pump is 0.15 mL/min (1 mL of insulin solution contains 100 U of insulin), and normally 1 mg of glucagon is diluted in 1 mL of solution, the maximum glucagon flow rate in Eq. \eqref{eq:gluccagonlevel} is set to $u_G^U = 0.15$ mg/min.

The amount of insulin administered in a bolus to a patient with a basal glucose level \AS{lower than 150 mg/dL normally ranges between 0.12 and 0.2 U/kg \cite{lorenzi1984duration}.
As the body mass of the {\it in-silico} patient we consider is 78 kg, we set $\phi_I^{U} = 16$ U}  in Eq. \eqref{eq:phiI}.
The maximum total amount of glucagon administered in one shot to a patient who is in a hypoglycemic state is 1 mg, and a second identical shot can be administered after thirty minutes.
We thus choose the maximum total amount of glucagon used (as defined in Eq. \eqref{eq:phiG}) throughout the five hour therapy to be $\phi_G^{U} = 1 $mg.

The choice of the initial condition $\bar{\textbf{x}}$ in Eq.\ \eqref{eq:initialcondition} is critical.
We select the initial condition so that the solution of our optimal control problem only attempts to regulate glucose in response to a meal.
In the results presented we have set the initial condition equal to the values of the states when $u_I(t) = u_b$ after a period of fasting (the final point of the blue curve in Fig.\ \ref{fig:constinput_1}{\it B}).
If we were to select any alternative initial condition then the solution to the optimal control problem would try to `correct' the initial condition as well, making comparisons between solutions difficult.
%
%

Once the parameters, bounds, the control time period and the initial condition are set, we solve the nonlinear optimal control problem using $\mathcal{PSOPT}$.
We first solve the optimal control problem without glucagon (\ie $u^U_G = 0$), and then we solve the optimal control problem using both insulin and glucagon.

To evaluate the effectiveness of the obtained results, we introduce the following measures.
\begin{itemize}
    \item The cumulative insulin $r_I(t)$ and cumulative glucagon $r_G(t)$ used up to time $t$,
    \begin{equation*}
	  r_I(t)  = \int_{t_0}^{t} u_I(\tau) d\tau, \qquad
	  r_G(t)  = \int_{t_0}^{t} u_G(\tau) d\tau.
    \end{equation*}
    \item The total amount of insulin $\phi_I=r_I(t_f)$ and the total amount of glucagon $\phi_G=r_G(t_f)$ used up to final time $t_f$.
    \item The integral of $BGI$ over the entire time period $[t_0,t_f]$,
    \begin{equation*}
        \Delta = \int_{t_0}^{t_f} BGI(G(t)) dt.
    \end{equation*}
    where a large $\Delta$ indicates that the patient is at higher risk of either hyperglycemia or hypoglycemia for a prolonged period of time.
    \item  The maximum and minimum values attained by the blood glucose level over the entire time period $[t_0,t_f]$,
    \begin{equation*}
    G^{\max} = \max_{t \in [t_0,t_f]} G(t),\qquad
    G^{\min} = \min_{t \in [t_0,t_f]} G(t),
     \end{equation*}
    which measure the risk for either hyperglicemia or an hypoglicemia \cite{cox1994frequency,kovatchev2005quantifying}, respectively.
\end{itemize}

\subsection{Insulin as Control Input} 

In this section we use only insulin as control input, \ie we set $u_G = 0$ in Eq.\ \eqref{eq:ode_fx}.
As the orders of magnitude of the terms $BGI$ and $u_I^p$ in the objective function are different, it is important to find the appropriate values of the scaling factors $\alpha_p$ and $\alpha_I$.
In what follows, we use a {\it Pareto-front} analysis to determine these values.
We first rewrite the objective function as
\begin{equation}
  J = \int_{t_0}^{t_f}  \left[ \varepsilon BGI(G(t)) +  u_I^p \right] dt
\end{equation}
where $\varepsilon = {\alpha_p}/{\alpha_I}$. 
\begin{figure}[t!]
	\begin{adjustwidth}{- 1 in}{0in} 
		\includegraphics[width =1.15\textwidth]{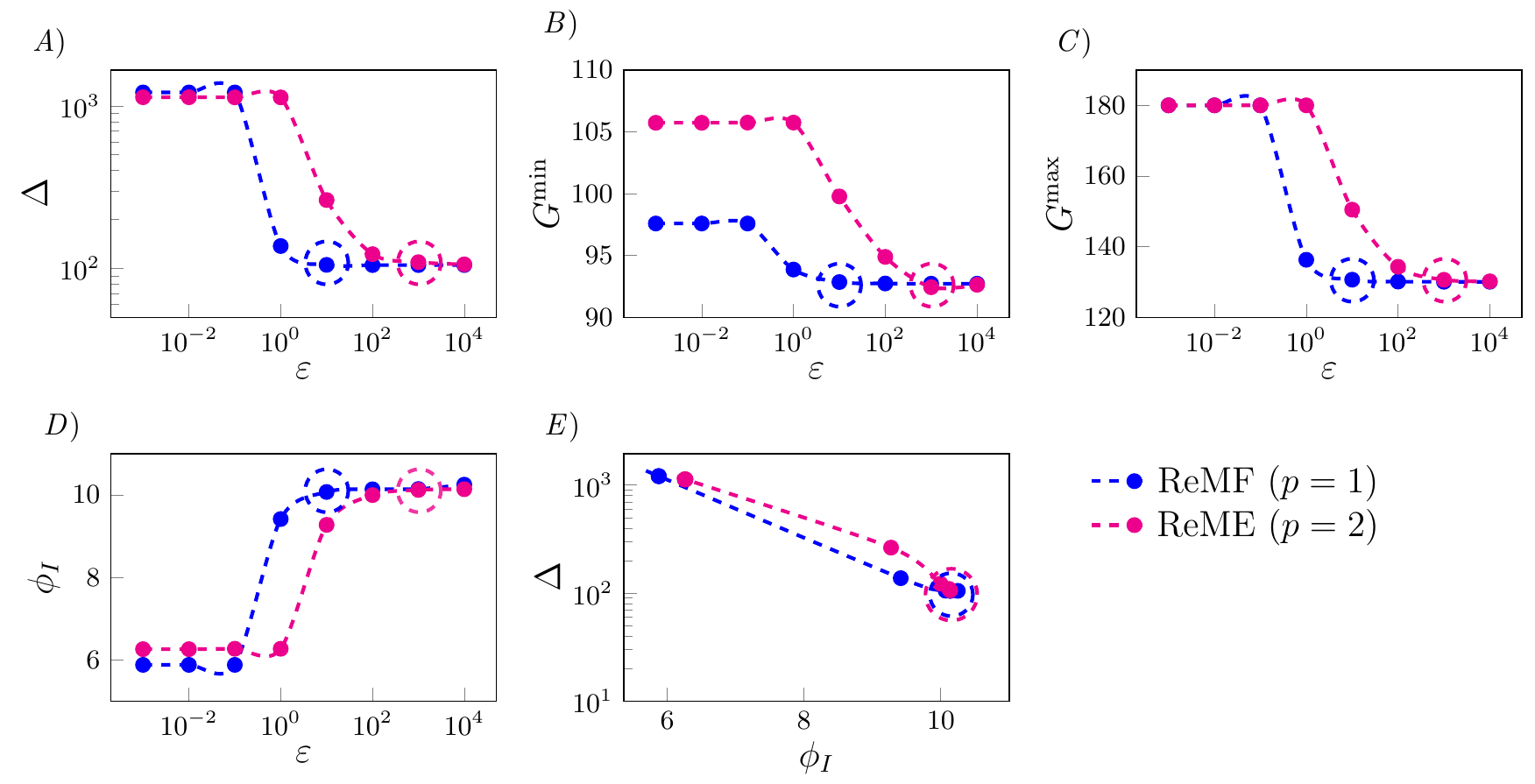}
		\caption{Performance of the optimal control solution as a function of $\varepsilon$. Large (small) values of $\varepsilon$ correspond to a large (small) weight associated with the $BGI$ index in the objective function, compared to the weight for insulin expenditure.
		The first four plots show our metrics as functions of the objective function coefficients: \textit{A}) $\Delta$ vs. $\varepsilon$, \textit{B}) $G^{\min}$ vs. $\varepsilon$, \textit{C}) $G^{\max}$ vs. $\varepsilon$, and \textit{D}) $\phi_I$ vs. $\varepsilon$.
		\textit{E}) We also project the Pareto front into the $\Delta$ - $\phi_I$ plane.
		We see a clear trade-off between $\Delta$ and $\phi_I$ as we vary $\varepsilon$.
		By increasing $\varepsilon$  we can decrease the values of $\Delta$ and $G^{\max}$.
		However, the values of $\Delta$ and $G^{\max}$ do not further decrease for $\varepsilon$ larger than $10$ for the ReMF problem ($p =1$) and the value of $\Delta$ does not further decrease for $\varepsilon$ larger than  $10^3$ for the ReME problem ($p = 2$).
		We choose $\varepsilon = 10$ for $p =1$ and $\varepsilon = 10^3$ for $p =2$, which are indicated by dashed circles in the figure, for the remaining simulations.} 
		\label{fig:pareto_front}
	\end{adjustwidth}
\end{figure}
In Figs.\ \ref{fig:pareto_front}({\it A--D}) we plot $\Delta$, $G^{\min}$, $G^{\max}$ and $\phi_I$ as functions of the coefficient $\varepsilon$. 
By looking at these plots, we see that the four measures can be divided into two groups.
\AS{On the one hand, $\Delta$ and $G^{\max}$ (panels {\it A} and {\it C}), improve (decrease) as $\varepsilon$ increases, with a sharp transition around $\varepsilon=10$ for the ReMF problem and around $\varepsilon=10^3$ for the ReME problem. On the other hand, $G^{\min}$ and $\phi_I$ (panels {\it B} and {\it D}), behave in the opposite way, \ie they improve (insulin decreases and  the minimum glucose level increases) as $\varepsilon$ decreases, again with a sharp transition around $\varepsilon=10$ for the ReMF problem and around $\varepsilon=10^3$ for the ReME problem. 
Because the four curves in Figs.\ \ref{fig:pareto_front}({\it A--D}) are monotone, all the points are Pareto-efficient, \ie it is not possible to improve one objective (\eg $\Delta$) without worsening the other one (\eg $\phi_I$). 
We notice that past a certain value of $\varepsilon$ ($10$ in the ReMF case, $10^3$ in the ReME case) $\Delta$ and $G^{\max}$ do not further decrease and $G^{\min}$ and $\phi_I$ remain unchanged. 
We choose as weights $\alpha_p = 10$ and $\alpha_I = 1$ for $p =1$, while we choose $\alpha_p = 10^3$ and $\alpha_I = 1$ for $p =2$ (these are highlighted by dashed circles in Fig.\ \ref{fig:pareto_front}).
The reason for these choices (for both values of $p$)  is that these values yield $\phi_I\sim 10$ units, which is equal to two thirds of the maximum amount of insulin that can be supplied ($\phi_I^U$), and $G^{\min}\sim 93$mg/dL, which is far from the hypoglycemic risk region.}

%

In Fig.\ \ref{fig:pareto_front}{\it E}, we plot a projection of the Pareto front in the $\Delta$ and $\phi_I$ plane.
Looking at this plot, the trade-off between $\Delta$ and $\phi_I$ is evident; if the total amount of insulin expenditure increases, $\Delta$ decreases and vice-versa. 
The ReMF and the ReME therapies can also be compared in Fig.\ \ref{fig:pareto_front}{\it E}.
The ReMF Pareto front dominates the ReME one (both $\Delta$ and $\phi_I$ are lower on the blue curve ($p=1$) compared to the magenta curve ($p=2$)).
This indicates that a shot of insulin (the optimal solution of a ReMF problem is typically a pulsatile function) performs slightly better in terms of $\Delta$ than a therapy in which the drug is delivered over a longer period of time while using less insulin. 

\begin{figure}[b!]
	\begin{adjustwidth}{- 1 in}{0in} 
	\includegraphics[width=1.15\textwidth]{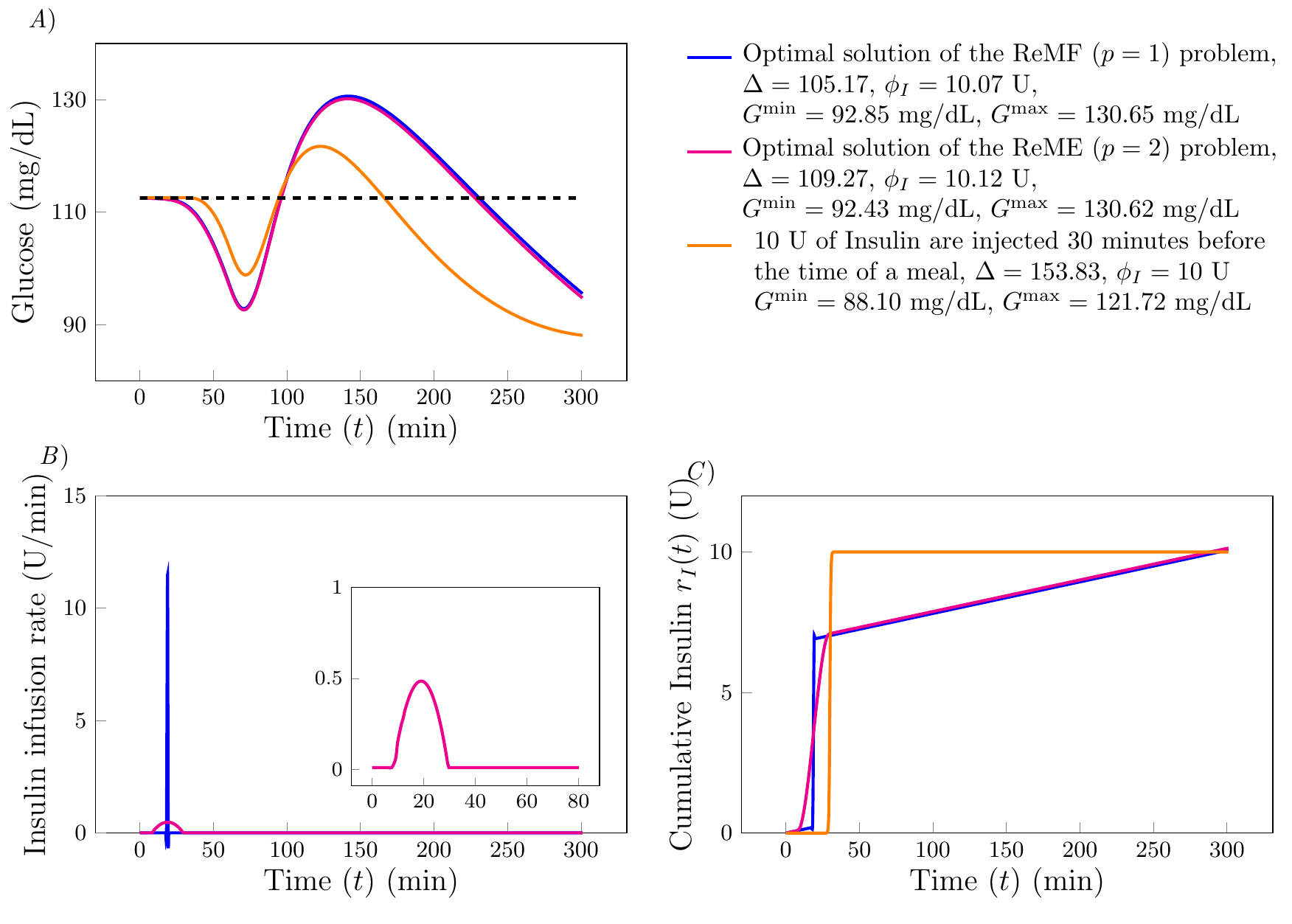}
	\caption{\textit{A}) The time evolution of glucose $G(t)$ (in mg/dL).  The blue curve corresponds to the pulsatile optimal insulin supply rate $u_I(t)$ (shown in \textit{B}) obtained by solving the ReMF problem. The magenta curve corresponds to the continuous optimal insulin supply rate $u_I(t)$ (shown in \textit{B}) obtained by solving the ReME problem. The orange curve is the time evolution of $G(t)$ corresponding to the standard therapy (10 U of insulin injected 30 minutes before the time of the meal). \textit{B}) Time evolution of the optimal insulin infusion rates  $u_I(t)$ (in U/min). Color code is consistent with \textit{A}.  \textit{C}) Cumulative insulin supply $r_I(t)$ (in U) as a function of $t$.}
	\label{fig:mono}
	\end{adjustwidth}
\end{figure}
Figure \ref{fig:mono} shows the results of the optimal control problem for the selected values of $\alpha_p$ and $\alpha_I$.
The blue and magenta curves are the optimal solutions of the ReMF and of the ReME problem, respectively.
The orange curve corresponds to the case that 10 U of insulin are injected 30 minutes before the time of the meal, \ie, the standard therapy. 


\AS{We observe that for $p = 1$ the optimal insulin infusion rate is pulsatile with a pulse appearing at $t \sim 20$ minutes, which is 40 minutes before the time of the meal. We obtained qualitatively similar results for different choices of the model parameters, with the pulse typically appearing at a time in the interval $t \in [20,30]$ minutes.}
It is noteworthy that the optimal solution is close to the standard insulin based therapy for glucose regulation in diabetics.  
The optimal insulin infusion rate is continuous when we solve the ReME problem, also shown in the inset of Fig.\ \ref{fig:mono}{\it B}.
Note that the ReMF and ReME therapies perform very similarly with respect to glucose as the peak insulin infusion rate occurs at approximately the same time and the total amount of insulin administered is nearly equal.

\subsection{Insulin and Glucagon as Control Inputs} 
In the previous section we tuned the weights $\alpha_p$ and $\alpha_I$ inside the objective function \eqref{eq:obj}.
We now consider the case that $u_G > 0$ and we tune $\alpha_G$, the weight associated with the glucagon expenditure in the objective function \eqref{eq:obj}, by keeping $\alpha_p = 10$, $\alpha_I = 1$ for $p = 1$ and $\alpha_p = 10^3$, $\alpha_I = 1$ for $p = 2$ , as previously determined.
%
%
%

%
\begin{figure}[t!]
	\begin{adjustwidth}{- 1 in}{0in} 
	    \includegraphics[width =1.15\textwidth]{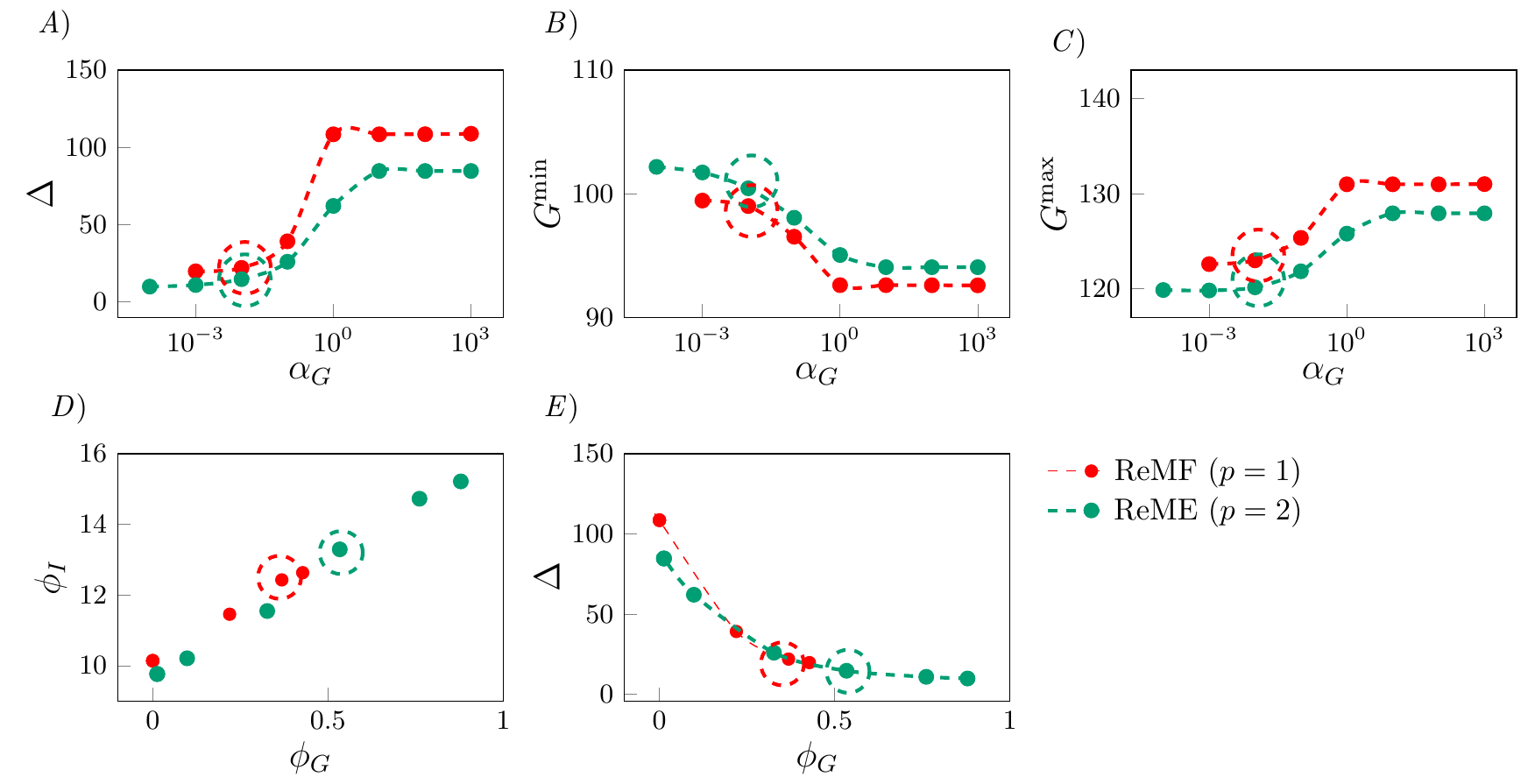}
		\caption{Performance of the optimal control solution as a function of $\alpha_G$. \textit{A}) $\Delta$ vs. $\alpha_G$.  \textit{B}) $G_{\min}$ vs. $\alpha_G$. \textit{C}) $G_{\max}$ vs. $\alpha_G$. \textit{D}) $\phi_I$ vs. $\phi_G$. \textit{E}) $\Delta$ vs. $\phi_G$. We select $\alpha_G = 10^{-2}$ for both of the REMF and REME problems, which are indicated by dashed circles in the figure. }
		\label{fig:pareto_front_dual}
	\end{adjustwidth}
\end{figure}
In Fig.\ \ref{fig:pareto_front_dual}{\it A}, \ref{fig:pareto_front_dual}{\it B} and \ref{fig:pareto_front_dual}{\it C},  we plot the optimal $\Delta$, $G^{\min}$ and $G^{\max}$ as functions of the parameter $\alpha_G$, respectively.
A large value of $\alpha_G$ indicates that we are placing a large weight on the expenditure of glucagon within the objective function \eqref{eq:obj}, \ie, the larger the value of $\alpha_G$, the less glucagon we use.
By looking at Fig.\ \ref{fig:pareto_front_dual}{\it A}, we observe that the values of $\Delta$ decrease as $\alpha_G$ decreases, \ie we can obtain lower (improved) values of $\Delta$ if we allow for a larger expenditure of glucagon.
\AS{We note that past a certain value of $\alpha_G$ ($10^{-2}$ in the both the ReMF and ReME problems) no further reduction in $\Delta$ is observed. }
As in the previous case, the maximum glucose level $G^{\max}$, shown in Fig.\ \ref{fig:pareto_front_dual}{\it C}, improves (decreases) when $\Delta$ improves (decreases). 
Interestingly, different from the previous case, also the minimum glucose level $G^{\min}$ (Fig. \ref{fig:pareto_front_dual}{\it C}) improves (increases) with $\Delta$ and $G^{\max}$: this is a consequence of the fact that we are using both insulin and glucagon as control inputs, which enables us to avoid both hypoglycemia and hyperglycemia. 

In  Fig.\ \ref{fig:pareto_front_dual}{\it D} we plot the projection of the Pareto front in the $(\phi_I,\phi_G)$ plane. 
\AS{By looking at the figure, $\phi_I$ and $\phi_G$ appear to be positively correlated and related by an approximately linear relation. While the timing of administration of insulin and glucagon is different, we see that overall the more insulin is used in the optimal solution, the more glucagon is used as well. This is because the two hormones have opposite effects in the regulation problem and thus they work so as to balance each other.  This is also consistent with the observation that with the dual drug therapy (insulin and glucagon) it becomes possible to simultaneously improve $\Delta$, $G^{\min}$, and $G^{\max}$.  
From the data in Fig.\ \ref{fig:pareto_front_dual}{\it D} we derive the following approximate linear relationship between $\phi_G$ and $\phi_I$,  
\begin{equation}\label{eq:linfunc}
  \begin{aligned}
    \AS{\phi_G(\phi_I) = 0.1596 \phi_I - 1.5796}
  \end{aligned}
\end{equation}
Obviously, glucagon should be used only when  $\phi_G(\phi_I)>0$.}

Panel \ref{fig:pareto_front_dual}{\it E} shows a projection of the Pareto front on the $(\phi_G,\Delta)$ plane.
We see again that the ReMF front dominates the ReME one, \ie a pulsatile therapy gives better results than a continuous therapy in terms of $\Delta$ and also uses lower amounts of the two drugs (smaller $\phi_G$, and thus smaller $\phi_I$ due to the positive correlation found in Fig. \ref{fig:pareto_front_dual}{\it D}).

The Pareto front is monotonically decreasing in Fig.\ \ref{fig:pareto_front_dual}{\it E} which indicates a trade-off between the total amount of drugs used and the achievable glucose control performance.
\AS{We choose the value of $\alpha_G$ for which the ratio between the increase in $\Delta$ and the decrease in $\phi_G$ is minimized, \ie $\alpha_G = 10^{-2}$ for both  ReMF and ReME problems, which are indicated by  dashed circles in the figure.}

\begin{figure}[t!]
	\begin{adjustwidth}{- 1 in}{0in} 
	 \includegraphics[width=1.15\textwidth]{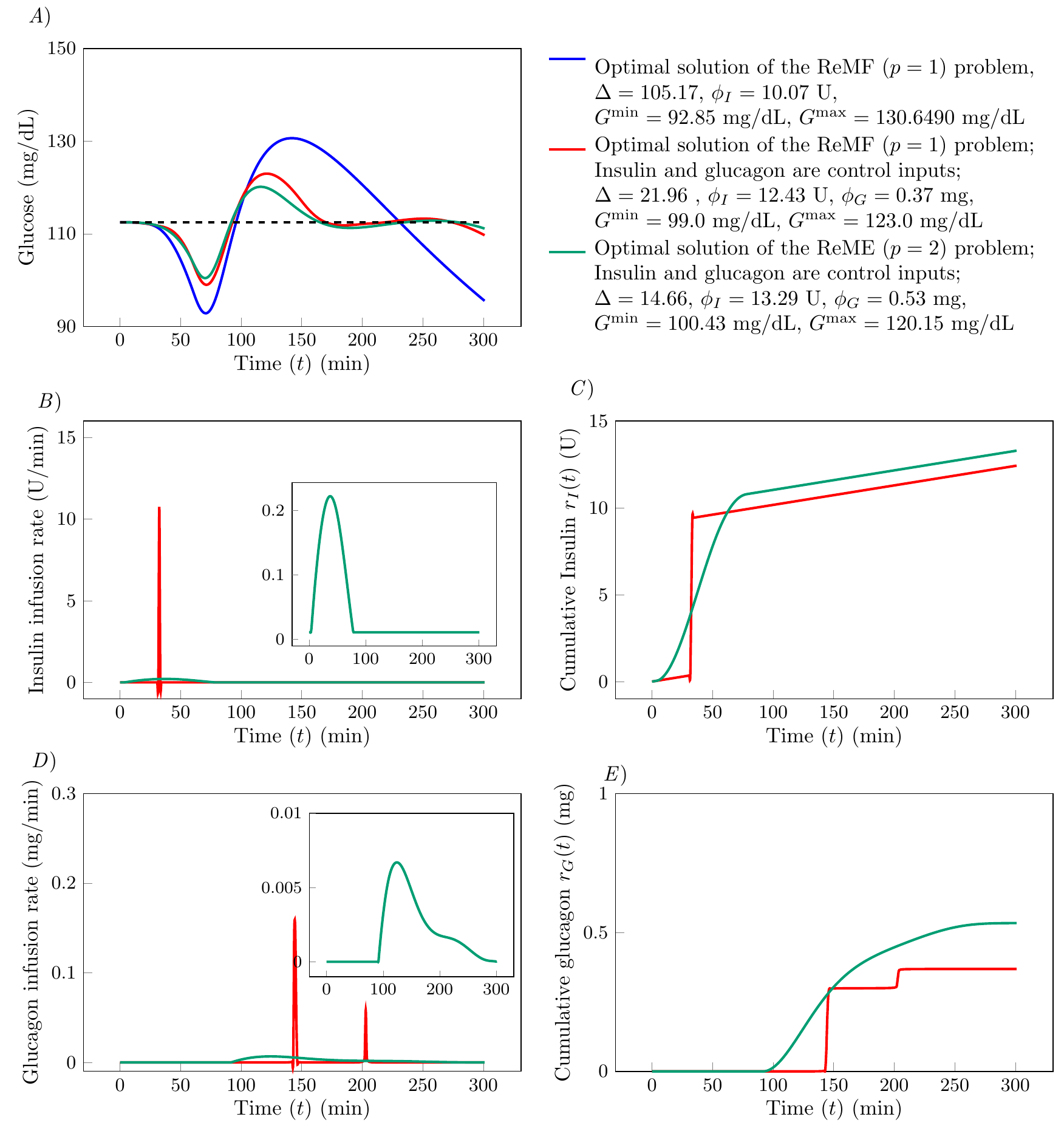}
		\caption{A) Time evolution of glucose $G(t)$ (in mg/dL). The blue curve corresponds to $u_I(t)$ obtained by solving the ReMF problem . The red curve corresponds to $u_I(t)$ and $u_G(t)$ obtained by solving the ReMF problem using the dual drug therapy. The green curve corresponds to $u_I(t)$ and $u_G(t)$ obtained by solving the ReME problem using the two-drug-therapy. \textit{B}) Time evolution of the insulin infusion rate  $u_I(t)$ (in mg/dL).  Color code is consistent with \textit{A}. \textit{C}) The cumulative insulin supply $r_I(t)$ as a function of time $t$. \textit{D}) Time evolution of the glucagon infusion rate  $u_G(t)$ (in mg/dL). \textit{E}) The cumulative glucose supply $r_G(t)$ as a function of time $t$. }
		\label{fig:dual}
	\end{adjustwidth}
\end{figure}
\indent
\AS{Figures \ref{fig:dual}{\it A} and \ref{fig:dual}{\it B} show the results of the optimal control problem for $\alpha_p= 10$, $\alpha_I=1$ and $\alpha_G = 10^{-2}$ when $p = 1$; and $\alpha_p= 10^3$, $\alpha_I=1$ and $\alpha_G = 10^{-2}$ when $p = 2$.  }
In Fig.\ \ref{fig:dual}{\it A} we plot the time evolution of glucose $G(t)$ for the different optimal solutions.
The blue curve corresponds to the solution of the ReMF problem when only insulin is used (the blue curve in Fig. \ref{fig:mono}{\it A}).
The red and green curves correspond to the solution of the ReMF and the ReME problems for the dual drug therapy.
We observe that $G(t)$ reaches the desired level $G_d$ faster if we use both insulin and glucagon as control inputs, compared to the case that only insulin is used.
We also see that in this case both $G^{\max}$ decreases  and  $G^{\min}$ increases.  
We therefore conclude that the therapy with both insulin and glucagon performs better than the therapy with only insulin, as the risks for both hypoglycemia and hyperglycemia are reduced and glucose fluctuations are suppressed. 

In Fig.\ \ref{fig:dual}{\it B} we plot the optimal insulin infusion rates and in Fig.\ \ref{fig:dual}{\it C} we plot the cumulative insulin supply $r_I(t)$ as a function of time $t$.
\AS{We observe that for the ReMF problem, the pulse in insulin appears at $t = 32$ minutes in the case that both insulin and glucagon are used ($28$ minutes before the meal), whereas the pulse appears at $t = 20$ minutes when only insulin is used.   
From Fig.\ \ref{fig:dual}{\it D}, we see that, for the ReMF problem, the glucagon delivery function is pulsatile with a main pulse appearing at $t = 145$ min (one hour and 25 minutes after the meal) and a secondary pulse appearing at $t = 203$.}
\AS{The dual drug therapy shows a noticeable difference between the ReMF solution and the ReME solution.
As expected, the solutions of the ReME problem are continuous.
The glucose response to the ReME therapy is better than the glucose response to the ReMF solution. Specifically, the green curve has smaller oscillations (in panel {\it A}) at the cost of small increases in the total amounts of used insulin and glucagon (compare panels {\it C} and {\it E})).}


Based on the results in Fig.\ \ref{fig:dual}, we propose a possible {\it ad-hoc} dual drug therapy to be used as an alternative to the standard therapy.
\AS{Rather than administering insulin half an hour before the meal (standard therapy), better glucose regulation can be achieved with a slightly larger insulin injection half an hour before a meal followed by a glucagon injection one hour and thirty minutes after a meal.}
\AS{The insulin injection of the \textit{ad-hoc} dual drug therapy is 25\% larger than the one used in the standard therapy, which is consistent with the relation between $\phi_I$ for the monotherapy ReMF optimal solution and the one used in the dual drug therapy.}

%

\begin{figure}[t!]
	\begin{adjustwidth}{- 1 in}{0in} 
		\includegraphics[width =1.15\textwidth]{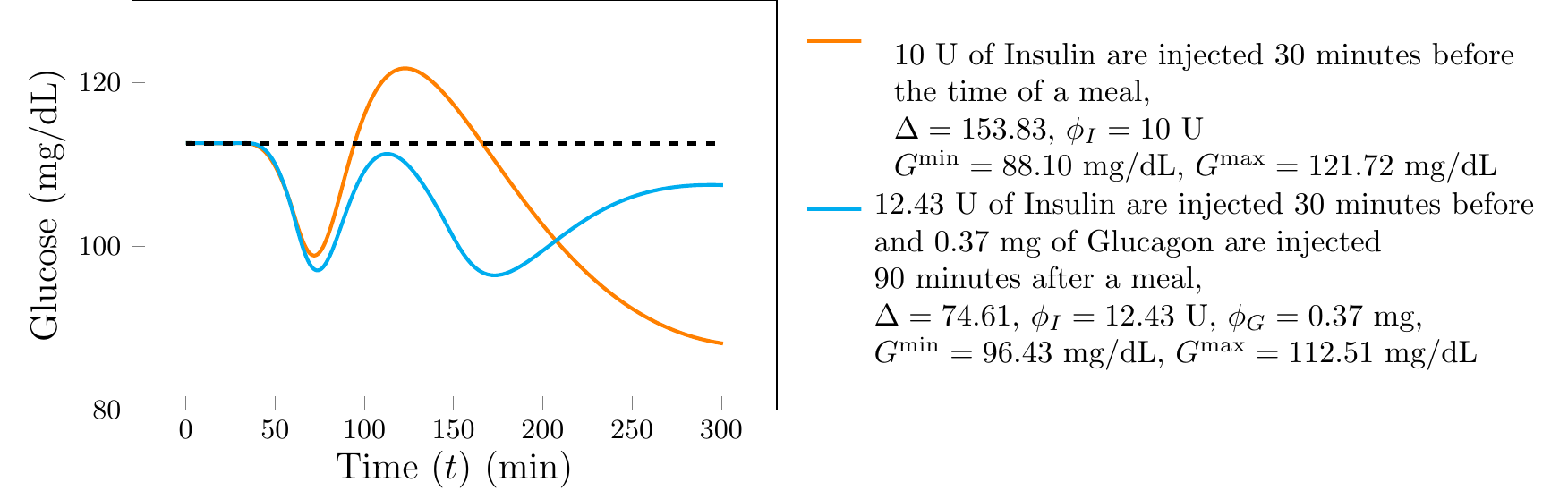}
		\caption{Comparison between the glucose response to the standard insulin base therapy (orange curve) and the proposed \textit{ad-hoc} dual therapy (cyan curve).}
		\label{fig:ruleG}
	\end{adjustwidth}
\end{figure}
\AS{In Fig. \ref{fig:ruleG} we present a comparison between the glucose response to the standard insulin base therapy (orange curve) and the proposed \textit{ad-hoc} dual therapy (cyan curve)  for the case of a meal with $70$ grams of glucose  (for the particular patient considered this corresponds to $10$ units of insulin half an hour before the meal) and the proposed \textit{ad-hoc} dual drug therapy (which consists of $12.43$ units of insulin thirty minutes before the meal and $0.40$ mg of glucagon one hour and thirty minutes after the meal).}
We observe that the \textit{ad-hoc} dual drug therapy performs better in terms of all of the proposed measures ($\Delta$, $G^{\min}$, $G^{\max}$, $\phi_I$ and $\phi_G$) as opposed to the standard insulin based therapy.

\subsection{Robustness Analysis}
\AS{We now analyze the robustness of the optimal control therapies we have proposed with respect to model parameter mismatches, which is a fundamental step for implementation of model based control.
We consider two different types of mismatches. The first type accounts for variability in the patient's behavior, in terms of both the time of the meal $\tau_D$ and the amount of glucose intake $D$.
The second type accounts for deviations in the parameter estimation, as well as the temporal variability of the parameters that a patient may experience during the day \cite{visentin2018uva}.}

\subsubsection{Robustness Against Variability of the Meal Time and Glucose Intake}
\AS{In this section we analyze the robustness of the optimal ReMF therapies (both monotherapy and dual therapy)  with respect to the two ``control" parameters the patient has. 
The first one is the variation in the meal time, ($\bar{\tau}_D - \tau_D$) (min), where $\bar{\tau}_D$ is the time of a meal and $\tau_D$ is the time of a meal we assumed in order to compute the optimal therapies. 
The second one is the variation of glucose in the meal, ($\bar{D} - D$), where $\bar{D}$ is the glucose intake in a meal and $D$ is the glucose intake we assumed to compute the optimal therapies. 
We consider variations in the meal time $\bar{\tau}_D$ in the interval $[30,90]$ min and  variations of the glucose intake $\bar{D}$ in the interval $[40,100]$ g.} 

\begin{figure}[b!]
	\begin{adjustwidth}{- 1 in}{0in} 
		\includegraphics[width =1.15\textwidth]{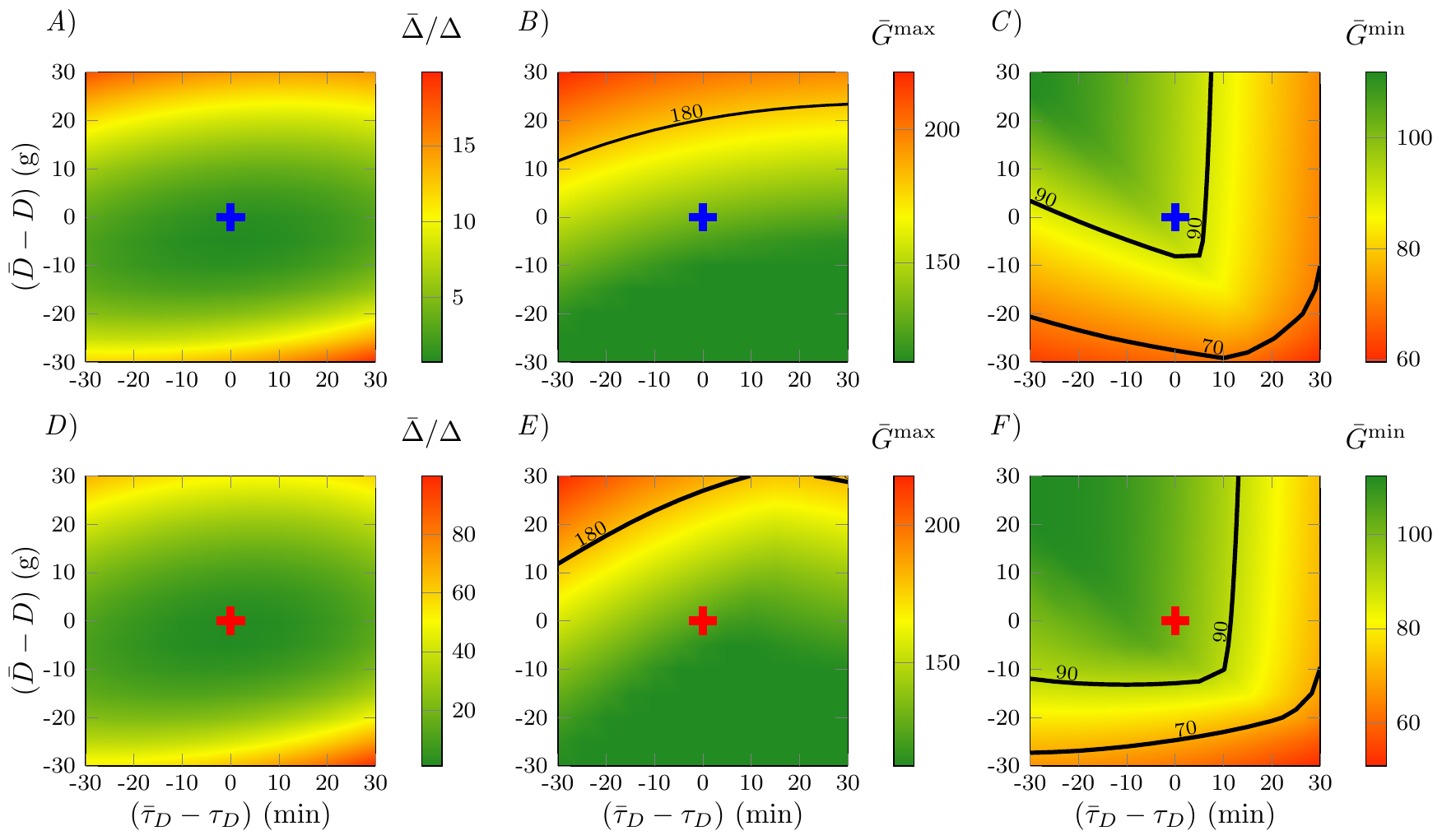}
		\caption{Robustness of the optimal control solution against variations in the meal timing and the amount of glucose in the meal.
		\textit{A})--\textit{C}) show the results obtained for the ReNF problem ($p = 1$) with only insulin provided, \textit{D})-\textit{F}) ReMF ($p = 1$) problem with both insulin and glucagon provided.
		Cross symbols indicate the application of the optimal control therapies for $\bar{D}=D$ and $\bar{\tau}_D=\tau_D$. The blue cross symbols correspond to the optimal therapies for the ReMF problem with only insulin. The red cross symbols correspond to the optimal therapies for the ReMF problem with both insulin and glucagon. 
		\textit{A)} and \textit{D)} are plots of  $\bar{\Delta}/\Delta$ in the control parameters space $(\bar{\tau}_D,\bar{D})$.
		\textit{B)} and \textit{E)} are the plots of $\bar{G}^{\max}$ in the control parameters space $(\bar{\tau}_D,\bar{D})$.
		\textit{C)} and \textit{F)} are the plots of $\bar{G}^{\min}$ in the control parameters space $(\bar{\tau}_D,\bar{D})$.
		}
		\label{fig:meal}
	\end{adjustwidth}
\end{figure}

\AS{The results of this study are illustrated in Fig.\ \ref{fig:meal}. Figure \ref{fig:meal} provides a visual assessment of the quality of the optimal therapies in terms of the three proposed measures $\Delta$, $G^{\max}$ and $G^{\min}$ (the over-bar stands for evaluation at the perturbed parameter values $(\bar\tau_D,\bar D)$). 
The color in Fig.\ \ref{fig:meal} varies according to the control performance from green (good) to red (dangerous).
In the upper panels (\textit{A}--\textit{C}) we consider the optimal ReMF monotherapy, while in the lower panels (\textit{D}-\textit{F}) we consider the optimal ReMF dual thearpy.
Cross symbols indicate the application of the optimal control therapies under ideal condition, \ie when $\bar\tau_D = {\tau}_D$ and $\bar{D} = D$. 
The black curves labeled by 180, 90 and 70 in Figs.\ \ref{fig:meal}\textit{B},  \ref{fig:meal}\textit{C}, \ref{fig:meal}\textit{D}, \ref{fig:meal}\textit{E}  are the curve level plots for  $\bar{G}^{\max} = G^U$, $\bar{G}^{\min} = G^L$ and $\bar{G}^{\min} = 70$, respectively. The black curves labeled by 180 in Figs.\ \ref{fig:meal}\textit{B}, and \ref{fig:meal}\textit{E}, are the curve level plots for  $\bar{G}^{\max} = G^U$.} 

\AS{We see from Figs. \ref{fig:meal}\textit{A} and \ref{fig:meal}\textit{D} that the optimal therapies for the ReMF problem (using only insulin or both insulin and glucagon) are robust with respect to variations in the control parameters: $\bar \Delta$ remains well bounded in most of the considered parameter space. 
In particular we see from Figs. \ref{fig:meal}\textit{B} and \ref{fig:meal}\textit{E} that the proposed optimal therapies are robust against hyperglicemic events: for example, even if $\bar{D}$ exceeds $D$ by 50\% and $\bar{\tau}_D$ exceeds $\tau_D$ by 30 minutes, the patient will not enter the hyperglycemic regime ($G^{\max}>300$).  
Figures \ref{fig:meal}\textit{C} and \ref{fig:meal}\textit{F} reveal that the proposed therapies suffer from a certain lack of robustness with respect to hypoglycemic events ($G^{\min}<70$), the most dangerous ones. The dangerous cases are, however, confined to extreme situations in which $\bar{D} < 0.5 D$ and $\bar{\tau}_D=\tau_D +30$ minutes. 
Figures \ref{fig:meal}\textit{C} and \ref{fig:meal}\textit{F} show also that the optimal therapy for the ReMF problem with both insulin and glucagon is more robust 
(larger green region and smaller yellow region) 
than the  optimal therapy for ReMF problem with only insulin (smaller green region and larger yellow region):
thus the use of glucagon alleviates the risk of severe, life-threatening hypoglycemia.
}

\AS{
We obtain qualitatively similar results when performing the same analysis for the other therapies we proposed (the ReME therapies and the \textit{ad-hoc} dual drug therapy).
}

\subsubsection{Robustness to Parameter Mismatches}
\begin{figure}[b!]
	\begin{adjustwidth}{- 1 in}{0in} 
		\includegraphics[width =1.15\textwidth]{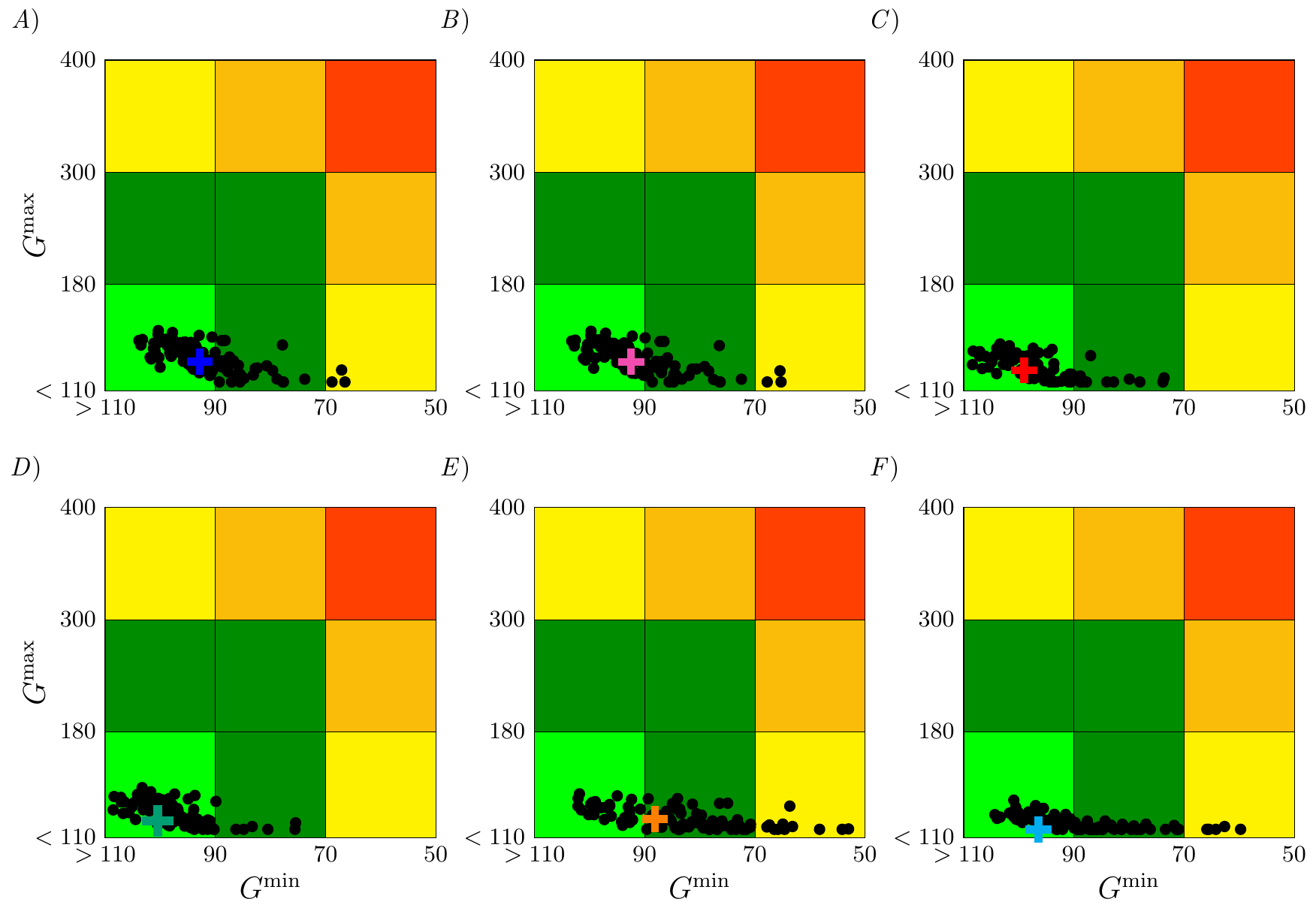}
		\caption{Robustness of the optimal control solution against parameter perturbations of the system and CVGA in the $G^{min}, G^{max}$ plane.
		The analysis is performed for \textit{A}) ReMF ($p = 1$) problem with only insulin provided, \textit{B}) ReME ($p = 2$) problem with only insulin provided, \textit{C}) ReMF ($p = 1$) problem with both insulin and glucagon provided, \textit{D}) ReME ($p = 2$) problem with both insulin and glucagon provided, \textit{E}) the standard therapy, and \textit{F}) the proposed \textit{ad-hoc} dual drug therapy.
		Cross symbols indicate the application of the optimal control therapies to the unperturbed systems.}
		\label{fig:cvga}
	\end{adjustwidth}
\end{figure}

\AS{
We consider perturbation of the model parameters up to $20\%$ of their nominal values,
\begin{equation}
  \bar{\Theta}_i = \Theta_i(1+\varphi),  
\end{equation}
where $\varphi$ is a random number from a normal distribution $\mathcal{N}(0, 0.067^2)$, $\Theta_i$ is a nominal parameter for a given patient with basal glucose level $G_b$ and  $\bar{\Theta}_i$ represents the associated perturbed parameter.}
We then apply the optimal insulin and glucagon dosing, calculated for the unperturbed system, to 100 perturbed systems.
\AS{This is analogous to testing the computed optimal control therapy on a specific  patient, but the patient's parameters may vary due to imperfect knowledge or due to the parameter variability throughout the day.}
The results of this study are illustrated with a Control Variability Grid Analysis (CVGA), see Fig.\ \ref{fig:cvga}.
The CVGA provides a simultaneous visual and numerical assessment of the overall performance of the glycemic control strategies in terms of the achieved minimum/maximum glucose values in the space of parameters mismatches.
In Fig.\ \ref{fig:cvga}, points in the light green region indicate accurate blood glucose control while points in the dark green regions indicate the patient is not immediately at risk of either hypoglycemia or hyperglycemia.
Points in the top two yellow/orange regions indicate an elevated risk of hyperglycemia and points in the the right two yellow/orange regions indicate an elevated risk of hypoglycemia.
Finally, points in the red corner region indicate an elevated risk of both hyperglycemia and hypoglycemia.
Each point reported in the figure is a plot of $G^{\max}$ vs. $G^{\min}$.
\AS{Here, the black dots correspond to the glucose response when a certain therapy is applied to a system with perturbed parameters.}
Cross symbols indicate application of the optimal control therapies to the unperturbed systems. \\
\indent
\AS{For the of monotherapy (ReMF in Fig. \ref{fig:cvga}{\it A} and ReME in Fig. \ref{fig:cvga}{\it B}) we find that the control is 67\% and 61\% accurate, respectively.
For the dual therapy case (ReMF in Fig. \ref{fig:cvga}{\it C} and ReME in Fig. \ref{fig:cvga}{\it D}) we find the control is more accurate than for the case of monotherapy, 92\% and 94\% accurate, respectively. 
The least robust control is obtained with the standard therapy (shown in Fig. \ref{fig:cvga}{\it E}), attaining only 37\% accuracy. 
Note that the optimal dual drug therapies (Figs. \ref{fig:cvga}\textit{C} and \ref{fig:cvga}\textit{D}) are not only  more robust than the optimal insulin therapies (Figs. \ref{fig:cvga}\textit{A} and \ref{fig:cvga}\textit{B}), but also than the standard therapy (\ref{fig:cvga}\textit{E}). 
We also see that the \textit{ad hoc} therapy  (\ref{fig:cvga}\textit{E}) is more robust than the standard therapy (\ref{fig:cvga}\textit{E}). }
%

\section{Discussion}
\AS{In this paper we have used the Glucose-Insulin-Glucagon mathematical model proposed in \cite{dalla2007gim,man2014uva,visentin2018uva}, which describes how the body responds to exogenously supplied insulin and glucagon in patients affected by Type I diabetes and designed an optimal dosing schedule of either insulin or insulin and glucagon together to regulate the blood glucose index (BGI), while limiting the total amount of insulin and glucagon administered.
The numerical optimal control software $\mathcal{PSOPT}$ has been used to solve this optimal control problem.
While the numerical solution requires knowledge of the set of model parameters, which are patient specific, the solutions we obtain provide insight into the best possible glucose regulation with insulin or with insulin and glucagon together.}
\AS{Our approach is in agreement with the results of references \cite{townsend2017characterisation,townsend2018control,townsend2018optimality}, in which simplified models are used to analytically establish general theoretical properties and control limitations for the glucose regulation problem.}

\AS{Two distinct regulation problems have been considered: the minimum fuel problem (ReMF) which yields pulsatile (shot-like) type solutions and the minimum energy problem (ReME) which yields longer periods of time over which insulin is administered but with smaller delivery rates.
This has allowed us to compare standard therapies which typically consist in shots of insulin with therapies in which insulin is delivered continuously.}
\AS{In \cite{goodwin2015fundamental,goodwin2018fundamental} it has been proven that the optimal control is pulsatile when the aim of the control is to minimize the variation in the maximum and minimum output response, the system is positive (like the one we are considering) and the disturbance (the meal, in our case) is pulsatile. Our work indicates that pulsatile control is still a good choice when more complex objective functions are chosen. Moreover, a pulsatile control appears to be optimal for alternative more realistic models of the meal (for example, a meal that is consumed over a window of 15 minutes).} 
\AS{We also see that a continuous drug delivery can achieve better results in the case of the dual therapy, thus pointing out the importance of developing a commercial pump able to deliver both insulin and glucagon.}

\AS{For both the ReMF and ReME problems, we compute the optimal drug dosing schedules when only insulin is available and when both insulin and glucagon are available.
The solution of the insulin only ReMF problem, astoundingly, is nearly equal to the standard method of insulin based glucose regulation.
Similarly, the solution of the ReMF problem when insulin and glucagon are used is also pulsatile, except that the amount of insulin administered is larger and the administration time is closer to the time of the meal, while the glucagon is mostly delivered in a shot about an hour and thirty minutes after the meal.}

\AS{The solution for the ReME problem when insulin only is available as well as when both insulin and glucagon are available is different from the ReMF solution in that, rather than being pulsatile, insulin and possibly glucagon are delivered at a slower rate over longer periods of time.
Nonetheless, the total amount of insulin and possibly glucagon is about the same, and the peak of the longer delivery time occurs approximately at the same time of the shot according to the solution of the ReMF problem.
The obtained  glucose profiles for the optimal ReMF and ReME problem solutions do not differ too much from each other: taken together, these results indicate that the amounts of insulin and glucagon, and the peak times of delivery, are the most important factors to determine when computing the optimal solutions.}

\AS{Based on the above results, we have proposed the following \textit{ad hoc} therapy when insulin and glucagon are used in combination: \textit{Administer a shot of insulin (with $5\%$ more insulin than the amount required by the standard therapy based on the planned meal) $30$ minutes before eating. Administer a shot of glucagon of an amount specified by Eq.\ \eqref{eq:linfunc} one hour and thirty minutes after completing the meal}.
This therapy must be used with caution as the amount of insulin injected can lead to hypoglycemia if the shot of glucagon is not administered as well.}

\AS{All optimal dosing schedules we computed were tested for robustness with respect to variations in the meal timing and size and with respect to variability of the parameters.
The therapies we proposed typically maintain the patient in the healthy region even under variable conditions and patient behavior. 
Note, however, that the proposed therapies are open-loop (the drug schedule is computed only from the condition of the patient at the initial time), thus cannot  compensate for unexpected behavior that can arise due to modeling simplifications (\eg we do not consider how physical activity influences the blood glucose production and consumption \cite{hajizadeh2018incorporating,beneyto2018new}), measurement noise or bias. A step towards the real application of our methodology is a real-time closed-loop strategy; this is possible, since the typical time needed to compute an optimal solution \AS{on a standard laptop (i7-8550U CPU with 16GB RAM)} is around 2 minutes.} Another main limitation of our study is that  real life constraints and long term physiological effects may make a therapy based on exogenous administration of both insulin and glucagon impractical.

\AS{Our optimal control strategies require knowledge of the meal time and meal glucose amount.
This is somewhat undesirable, as recent advances in diabetes therapy have moved towards devices that do not require the user to provide information about the meals.
Our results emphasize the importance of knowing when the meals will occur, and that new dosing schedules would benefit from some knowledge about the meals. This seems to indicate that it would be beneficial to provide the pump with the ability to interpret the patient's behavior.}

\bibliography{diabetes}

\begin{thebibliography}{10}

\bibitem{dalla2007meal}
Dalla~Man C, Rizza RA, Cobelli C.
\newblock Meal simulation model of the glucose-insulin system.
\newblock IEEE Transactions on biomedical engineering. 2007;54(10):1740--1749.

\bibitem{dalla2007gim}
Dalla~Man C, Raimondo DM, Rizza RA, Cobelli C. GIM, simulation software of meal
  glucose—insulin model; 2007.

\bibitem{man2014uva}
Man CD, Micheletto F, Lv D, Breton M, Kovatchev B, Cobelli C.
\newblock The UVA/PADOVA type 1 diabetes simulator: new features.
\newblock Journal of diabetes science and technology. 2014;8(1):26--34.

\bibitem{visentin2018uva}
Visentin R, Campos-N{\'a}{\~n}ez E, Schiavon M, Lv D, Vettoretti M, Breton M,
  et~al.
\newblock The UVA/Padova Type 1 Diabetes Simulator Goes From Single Meal to
  Single Day.
\newblock Journal of diabetes science and technology. 2018;12(2):273--281.

\bibitem{steil2006feasibility}
Steil GM, Rebrin K, Darwin C, Hariri F, Saad MF.
\newblock Feasibility of automating insulin delivery for the treatment of type
  1 diabetes.
\newblock Diabetes. 2006;55(12):3344--3350.

\bibitem{magni2008evaluating}
Magni L, Raimondo DM, Man CD, Breton M, Patek S, De~Nicolao G, et~al.
\newblock Evaluating the efficacy of closed-loop glucose regulation via
  control-variability grid analysis.
\newblock Journal of Diabetes Science and Technology. 2008;2(4):630--635.

\bibitem{magni2009model}
Magni L, Raimondo DM, Dalla~Man C, De~Nicolao G, Kovatchev B, Cobelli C.
\newblock Model predictive control of glucose concentration in type I diabetic
  patients: An in silico trial.
\newblock Biomedical Signal Processing and Control. 2009;4(4):338--346.

\bibitem{palerm2008run}
Palerm CC, Zisser H, Jovanovi{\v{c}} L, Doyle~III FJ.
\newblock A run-to-run control strategy to adjust basal insulin infusion rates
  in type 1 diabetes.
\newblock Journal of process control. 2008;18(3-4):258--265.

\bibitem{bergenstal2013threshold}
Bergenstal RM, Klonoff DC, Garg SK, Bode BW, Meredith M, Slover RH, et~al.
\newblock Threshold-based insulin-pump interruption for reduction of
  hypoglycemia.
\newblock New England Journal of Medicine. 2013;369(3):224--232.

\bibitem{van2014feasibility}
van Bon AC, Luijf YM, Koebrugge R, Koops R, Hoekstra JB, DeVries JH.
\newblock Feasibility of a portable bihormonal closed-loop system to control
  glucose excursions at home under free-living conditions for 48 hours.
\newblock Diabetes technology \& therapeutics. 2014;16(3):131--136.

\bibitem{nimri2014md}
Nimri R, Muller I, Atlas E, Miller S, Fogel A, Bratina N, et~al.
\newblock MD-Logic overnight control for 6 weeks of home use in patients with
  type 1 diabetes: randomized crossover trial.
\newblock Diabetes Care. 2014; p. DC\_140835.

\bibitem{capel2014artificial}
Capel I, Rigla M, Garc{\'\i}a-S{\'a}ez G, Rodr{\'\i}guez-Herrero A, Pons B,
  Sub{\'\i}as D, et~al.
\newblock Artificial pancreas using a personalized rule-based controller
  achieves overnight normoglycemia in patients with type 1 diabetes.
\newblock Diabetes technology \& therapeutics. 2014;16(3):172--179.

\bibitem{mauseth2015stress}
Mauseth R, Lord SM, Hirsch IB, Kircher RC, Matheson DP, Greenbaum CJ.
\newblock Stress testing of an artificial pancreas system with pizza and
  exercise leads to improvements in the system’s fuzzy logic controller.
\newblock Journal of diabetes science and technology. 2015;9(6):1253--1259.

\bibitem{reddy2014feasibility}
Reddy M, Herrero P, El~Sharkawy M, Pesl P, Jugnee N, Thomson H, et~al.
\newblock Feasibility study of a bio-inspired artificial pancreas in adults
  with type 1 diabetes.
\newblock Diabetes technology \& therapeutics. 2014;16(9):550--557.

\bibitem{parker1999model}
Parker RS, Doyle FJ, Peppas NA.
\newblock A model-based algorithm for blood glucose control in type I diabetic
  patients.
\newblock IEEE Transactions on biomedical engineering. 1999;46(2):148--157.

\bibitem{parker2001intravenous}
Parker RS, Doyle FJ, Peppas NA.
\newblock The intravenous route to blood glucose control.
\newblock IEEE Engineering in Medicine and Biology Magazine. 2001;20(1):65--73.

\bibitem{gillis2007glucose}
Gillis R, Palerm CC, Zisser H, Jovanovic L, Seborg DE, Doyle~III FJ. Glucose
  estimation and prediction through meal responses using ambulatory subject
  data for advisory mode model predictive control; 2007.

\bibitem{lynch2001estimation}
Lynch SM, Bequette BW.
\newblock Estimation-based model predictive control of blood glucose in type I
  diabetics: a simulation study.
\newblock In: Bioengineering Conference, 2001. Proceedings of the IEEE 27th
  Annual Northeast. IEEE; 2001. p. 79--80.

\bibitem{bruttomesso2009closed}
Bruttomesso D, Farret A, Costa S, Marescotti MC, Vettore M, Avogaro A, et~al..
  Closed-loop artificial pancreas using subcutaneous glucose sensing and
  insulin delivery and a model predictive control algorithm: preliminary
  studies in Padova and Montpellier; 2009.

\bibitem{marchetti2008improved}
Marchetti G, Barolo M, Jovanovic L, Zisser H, Seborg DE.
\newblock An improved PID switching control strategy for type 1 diabetes.
\newblock ieee transactions on biomedical engineering. 2008;55(3):857--865.

\bibitem{fisher1991semiclosed}
Fisher ME.
\newblock A semiclosed-loop algorithm for the control of blood glucose levels
  in diabetics.
\newblock IEEE transactions on biomedical engineering. 1991;38(1):57--61.

\bibitem{bergman1979quantitative}
Bergman RN, Ider YZ, Bowden CR, Cobelli C.
\newblock Quantitative estimation of insulin sensitivity.
\newblock American Journal of Physiology-Endocrinology And Metabolism.
  1979;236(6):E667.

\bibitem{bergman1981physiologic}
Bergman RN, Phillips LS, Cobelli C.
\newblock Physiologic evaluation of factors controlling glucose tolerance in
  man: measurement of insulin sensitivity and beta-cell glucose sensitivity
  from the response to intravenous glucose.
\newblock The Journal of clinical investigation. 1981;68(6):1456--1467.

\bibitem{bergman1985assessment}
Bergman RN, Finegood DT, Ader M.
\newblock Assessment of insulin sensitivity in vivo.
\newblock Endocrine reviews. 1985;6(1):45--86.

\bibitem{thabit2015home}
Thabit H, Tauschmann M, Allen JM, Leelarathna L, Hartnell S, Wilinska ME,
  et~al.
\newblock Home use of an artificial beta cell in type 1 diabetes.
\newblock New England Journal of Medicine. 2015;373(22):2129--2140.

\bibitem{zavitsanou2015silico}
Zavitsanou S, Mantalaris A, Georgiadis MC, Pistikopoulos EN.
\newblock In Silico Closed-Loop Control Validation Studies for Optimal Insulin
  Delivery in Type 1 Diabetes.
\newblock IEEE Trans Biomed Engineering. 2015;62(10):2369--2378.

\bibitem{hovorka2004nonlinear}
Hovorka R, Canonico V, Chassin LJ, Haueter U, Massi-Benedetti M, Federici MO,
  et~al.
\newblock Nonlinear model predictive control of glucose concentration in
  subjects with type 1 diabetes.
\newblock Physiological measurement. 2004;25(4):905.

\bibitem{bequette2012challenges}
Bequette BW.
\newblock Challenges and recent progress in the development of a closed-loop
  artificial pancreas.
\newblock Annual reviews in control. 2012;36(2):255--266.

\bibitem{copp2018simultaneous}
Copp DA, Gondhalekar R, Hespanha JP.
\newblock Simultaneous model predictive control and moving horizon estimation
  for blood glucose regulation in type 1 diabetes.
\newblock Optimal Control Applications and Methods. 2018;39(2):904--918.

\bibitem{mccrimmon2010hypoglycemia}
McCrimmon RJ, Sherwin RS.
\newblock Hypoglycemia in type 1 diabetes.
\newblock Diabetes. 2010;59(10):2333--2339.

\bibitem{castle2010novel}
Castle JR, Engle JM, El~Youssef J, Massoud RG, Yuen KC, Kagan R, et~al.
\newblock Novel use of glucagon in a closed-loop system for prevention of
  hypoglycemia in type 1 diabetes.
\newblock Diabetes care. 2010;33(6):1282--1287.

\bibitem{batora2015contribution}
B{\'a}tora V, T{\'a}rn{\'\i}k M, Murga{\v{s}} J, Schmidt S, N{\o}rgaard K,
  Poulsen NK, et~al.
\newblock The contribution of glucagon in an artificial pancreas for people
  with type 1 diabetes.
\newblock In: American Control Conference (ACC), 2015. IEEE; 2015. p.
  5097--5102.

\bibitem{el2007adaptive}
El-Khatib FH, Jiang J, Damiano ER.
\newblock Adaptive closed-loop control provides blood-glucose regulation using
  dual subcutaneous insulin and glucagon infusion in diabetic swine.
\newblock Journal of Diabetes Science and Technology. 2007;1(2):181--192.

\bibitem{el2010bihormonal}
El-Khatib FH, Russell SJ, Nathan DM, Sutherlin RG, Damiano ER.
\newblock A bihormonal closed-loop artificial pancreas for type 1 diabetes.
\newblock Science translational medicine. 2010;2(27):27ra27--27ra27.

\bibitem{russell2012blood}
Russell SJ, El-Khatib FH, Nathan DM, Magyar KL, Jiang J, Damiano ER.
\newblock Blood glucose control in type 1 diabetes with a bihormonal bionic
  endocrine pancreas.
\newblock Diabetes care. 2012; p. DC\_120071.

\bibitem{el2014autonomous}
El-Khatib FH, Russell SJ, Magyar KL, Sinha M, McKeon K, Nathan DM, et~al.
\newblock Autonomous and continuous adaptation of a bihormonal bionic pancreas
  in adults and adolescents with type 1 diabetes.
\newblock The Journal of Clinical Endocrinology \& Metabolism.
  2014;99(5):1701--1711.

\bibitem{russell2014outpatient}
Russell SJ, El-Khatib FH, Sinha M, Magyar KL, McKeon K, Goergen LG, et~al.
\newblock Outpatient glycemic control with a bionic pancreas in type 1
  diabetes.
\newblock New England Journal of Medicine. 2014;371(4):313--325.

\bibitem{russell2016day}
Russell SJ, Hillard MA, Balliro C, Magyar KL, Selagamsetty R, Sinha M, et~al.
\newblock Day and night glycaemic control with a bionic pancreas versus
  conventional insulin pump therapy in preadolescent children with type 1
  diabetes: a randomised crossover trial.
\newblock The lancet Diabetes \& endocrinology. 2016;4(3):233--243.

\bibitem{el2017home}
El-Khatib FH, Balliro C, Hillard MA, Magyar KL, Ekhlaspour L, Sinha M, et~al.
\newblock Home use of a bihormonal bionic pancreas versus insulin pump therapy
  in adults with type 1 diabetes: a multicentre randomised crossover trial.
\newblock The Lancet. 2017;389(10067):369--380.

\bibitem{herrero2017coordinated}
Herrero P, Bondia J, Oliver N, Georgiou P.
\newblock A coordinated control strategy for insulin and glucagon delivery in
  type 1 diabetes.
\newblock Computer methods in biomechanics and biomedical engineering.
  2017;20(13):1474--1482.

\bibitem{boiroux2018adaptive}
Boiroux D, B{\'a}tora V, Hagdrup M, Wendt SL, Poulsen NK, Madsen H, et~al.
\newblock Adaptive model predictive control for a dual-hormone artificial
  pancreas.
\newblock Journal of Process Control. 2018;68:105--117.

\bibitem{shirin2018prediction}
Shirin A, Klickstein I, Feng S, Lin YT, Hlavacek WS, Sorrentino F.
\newblock Prediction of Optimal Drug Schedules for Controlling Autophagy.
\newblock Accepted for publication in Scientific Reports.
  2019;doi:{10.1038/s41598-019-38763-9}.

\bibitem{shirin2017optimal}
Shirin A, Klickstein IS, Sorrentino F.
\newblock Optimal control of complex networks: Balancing accuracy and energy of
  the control action.
\newblock Chaos: An Interdisciplinary Journal of Nonlinear Science.
  2017;27(4):041103.

\bibitem{kovatchev1997symmetrization}
Kovatchev BP, Cox DJ, Gonder-Frederick LA, Clarke W.
\newblock Symmetrization of the blood glucose measurement scale and its
  applications.
\newblock Diabetes Care. 1997;20(11):1655--1658.

\bibitem{kovatchev2005quantifying}
Kovatchev BP, Clarke WL, Breton M, Brayman K, McCall A.
\newblock Quantifying temporal glucose variability in diabetes via continuous
  glucose monitoring: mathematical methods and clinical application.
\newblock Diabetes technology \& therapeutics. 2005;7(6):849--862.

\bibitem{diabetes1993effect}
Control D, Group CTR.
\newblock The effect of intensive treatment of diabetes on the development and
  progression of long-term complications in insulin-dependent diabetes
  mellitus.
\newblock New England journal of medicine. 1993;329(14):977--986.

\bibitem{kirk2012optimal}
Kirk DE.
\newblock Optimal control theory: an introduction.
\newblock Courier Corporation; 2012.

\bibitem{chachuat2007nonlinear}
Chachuat B.
\newblock Nonlinear and dynamic optimization: From theory to practice; 2007.

\bibitem{jdrfwebsite}
JDRF website: http://www.jdrf.org; 2018.

\bibitem{klickstein2017energy}
Klickstein I, Shirin A, Sorrentino F.
\newblock Energy scaling of targeted optimal control of complex networks.
\newblock Nature communications. 2017;8:15145.

\bibitem{rao2009survey}
Rao AV.
\newblock A survey of numerical methods for optimal control.
\newblock Advances in the Astronautical Sciences. 2009;135(1):497--528.

\bibitem{ross2012review}
Ross IM, Karpenko M.
\newblock A review of pseudospectral optimal control: From theory to flight.
\newblock Annual Reviews in Control. 2012;36(2):182--197.

\bibitem{becerra2010solving}
Becerra VM.
\newblock Solving complex optimal control problems at no cost with PSOPT.
\newblock In: Computer-Aided Control System Design (CACSD), 2010 IEEE
  International Symposium on. IEEE; 2010. p. 1391--1396.

\bibitem{nocedal2006numerical}
Nocedal J, Wright S.
\newblock Numerical optimization.
\newblock Springer, New York, USA, 2006. 2006;.

\bibitem{wachter2006implementation}
W{\"a}chter A, Biegler LT.
\newblock On the implementation of an interior-point filter line-search
  algorithm for large-scale nonlinear programming.
\newblock Mathematical programming. 2006;106(1):25--57.

\bibitem{russell2015insulin}
Russell-Jones D, Gall MA, Niemeyer M, Diamant M, Del~Prato S.
\newblock Insulin degludec results in lower rates of nocturnal hypoglycaemia
  and fasting plasma glucose vs. insulin glargine: a meta-analysis of seven
  clinical trials.
\newblock Nutrition, Metabolism and Cardiovascular Diseases.
  2015;25(10):898--905.

\bibitem{userguide}
MiniMed 670G System User Guide website:
  https://www.medtronicdiabetes.com/sites/user-guides/MiniMed; 2018.

\bibitem{lorenzi1984duration}
Lorenzi M, Bohannon N, Tsalikian E, Karam JH.
\newblock Duration of type I diabetes affects glucagon and glucose responses to
  insulin-induced hypoglycemia.
\newblock Western Journal of Medicine. 1984;141(4):467.

\bibitem{cox1994frequency}
Cox DJ, Kovatchev BP, Julian DM, Gonder-Frederick LA, Polonsky WH, Schlundt DG,
  et~al.
\newblock Frequency of severe hypoglycemia in insulin-dependent diabetes
  mellitus can be predicted from self-monitoring blood glucose data.
\newblock The Journal of Clinical Endocrinology \& Metabolism.
  1994;79(6):1659--1662.

\bibitem{townsend2017characterisation}
Townsend C, Seron MM, Goodwin GC.
\newblock Characterisation of optimal responses to pulse inputs in the Bergman
  minimal model.
\newblock IFAC-PapersOnLine. 2017;50(1):15163--15168.

\bibitem{townsend2018control}
Townsend C, Seron MM, Goodwin GC, King BR.
\newblock Control Limitations in Models of T1DM and the Robustness of Optimal
  Insulin Delivery.
\newblock Journal of diabetes science and technology. 2018;12(5):926--936.

\bibitem{townsend2018optimality}
Townsend C, Seron MM.
\newblock Optimality of unconstrained pulse inputs to the Bergman minimal
  model.
\newblock IEEE Control Systems Letters. 2018;2(1):79--84.

\bibitem{goodwin2015fundamental}
Goodwin GC, Medioli AM, Carrasco DS, King BR, Fu Y.
\newblock A fundamental control limitation for linear positive systems with
  application to Type 1 diabetes treatment.
\newblock Automatica. 2015;55:73--77.

\bibitem{goodwin2018fundamental}
Goodwin GC, Carrasco DS, Seron MM, Medioli AM.
\newblock A fundamental control performance limit for a class of positive
  nonlinear systems.
\newblock Automatica. 2018;95:14--22.

\bibitem{hajizadeh2018incorporating}
Hajizadeh I, Rashid M, Turksoy K, Samadi S, Feng J, Sevil M, et~al.
\newblock Incorporating Unannounced Meals and Exercise in Adaptive Learning of
  Personalized Models for Multivariable Artificial Pancreas Systems.
\newblock Journal of diabetes science and technology. 2018;12(5):953--966.

\bibitem{beneyto2018new}
Beneyto A, Bertachi A, Bondia J, Vehi J.
\newblock A New Blood Glucose Control Scheme for Unannounced Exercise in Type 1
  Diabetic Subjects.
\newblock IEEE Transactions on Control Systems Technology. 2018;.

\bibitem{DMemail}
Man CD. Parameters; 2018.
\newblock private communication by email.

\bibitem{T1DMSEG}
The Epsilon Group; 2018.
\newblock Available from \url{https://tegvirginia.com/software/t1dms/}.

\bibitem{DM2014}
The implementation of the UVA/Pavoda model (1014) in this paper; 2019.
\newblock Available from \url{https://github.com/iklick/dallaman_2014}.

\bibitem{liu2011controllability}
Liu YY, Slotine JJ, Barab{\'a}si AL.
\newblock Controllability of complex networks.
\newblock Nature. 2011;473(7346):167--173.

\bibitem{ruths2014control}
Ruths J, Ruths D.
\newblock Control profiles of complex networks.
\newblock Science. 2014;343(6177):1373 -- 1376.

\bibitem{summers2014optimal}
Summers TH, Lygeros J.
\newblock Optimal sensor and actuator placement in complex dynamical networks.
\newblock IFAC Proceedings Volumes. 2014;47(3):3784--3789.

\bibitem{wang2012control}
Wang B, Gao L, Gao Y.
\newblock Control range: a controllability-based index for node significance in
  directed networks.
\newblock Journal of Statistical Mechanics: Theory and Experiment.
  2012;2012(04):P04011.

\bibitem{nepusz2012controlling}
Nepusz T, Vicsek T.
\newblock Controlling edge dynamics in complex networks.
\newblock Nature Physics. 2012;8(7):568--573.

\bibitem{yuan2013exact}
Yuan Z, Zhao C, Di Z, Wang WX, Lai YC.
\newblock Exact controllability of complex networks.
\newblock Nature communications. 2013;4(2447).

\bibitem{iudice2015structural}
Iudice FL, Garofalo F, Sorrentino F.
\newblock Structural permeability of complex networks to control signals.
\newblock Nature communications. 2015;6(8349).

\bibitem{gao2016control}
Gao XD, Wang WX, Lai YC.
\newblock Control efficacy of complex networks.
\newblock Scientific Reports. 2016;6(28037).

\bibitem{yan2015spectrum}
Yan G, Tsekenis G, Barzel B, Slotine JJ, Liu YY, Barab{\'a}si AL.
\newblock Spectrum of controlling and observing complex networks.
\newblock Nature Physics. 2015;11(9):779--786.

\bibitem{yan2012controlling}
Yan G, Ren J, Lai YC, Lai CH, Li B.
\newblock Controlling complex networks: how much energy is needed?
\newblock Physical review letters. 2012;108(21):218703.

\bibitem{sorrentino2007controllability}
Sorrentino F, di~Bernardo M, Garofalo F, Chen G.
\newblock Controllability of complex networks via pinning.
\newblock Physical Review E. 2007;75:046103.

\bibitem{gates2016control}
Gates AJ, Rocha LM.
\newblock Control of complex networks requires both structure and dynamics.
\newblock Scientific reports. 2016;6.

\bibitem{cornelius2013realistic}
Cornelius SP, Kath WL, Motter AE.
\newblock Realistic control of network dynamics.
\newblock Nature Communications. 2013;4:1942.
\newblock doi:{10.1038/ncomms2939}.

\bibitem{wang2016geometrical}
Wang LZ, Su RQ, Huang ZG, Wang X, Wang WX, Grebogi C, et~al.
\newblock A geometrical approach to control and controllability of nonlinear
  dynamical networks.
\newblock Nature Communications. 2016;7:11323.
\newblock doi:{10.1038/ncomms11323}.

\bibitem{zanudo2017structure}
Za{\~n}udo JGT, Yang G, Albert R.
\newblock Structure-based control of complex networks with nonlinear dynamics.
\newblock Proceedings of the National Academy of Sciences.
  2017;114(28):7234--7239.
\newblock doi:{10.1073/pnas.1617387114}.

\bibitem{klickstein2017locally}
Klickstein I, Shirin A, Sorrentino F.
\newblock Locally Optimal Control of Complex Networks.
\newblock Physical Review Letters. 2017;119(26):268301.
\newblock doi:{10.1103/PhysRevLett.119.268301}.

\end{thebibliography}
\bibliographystyle{unsrt}

\clearpage
\appendix

\renewcommand{\thesection}{S\arabic{section}}
\renewcommand{\thefigure}{S\arabic{figure}}
\renewcommand{\theequation}{S\arabic{equation}}
\renewcommand{\thetable}{S\arabic{table}}
\setcounter{figure}{0}
\setcounter{section}{0}
\setcounter{equation}{0}
\setcounter{table}{0}

\section*{Supplementary Information}

\section{GIGM Model and Parameters}\label{sec:model}

\subsection{Overview of GIGM Model with Type I Diabetics}

We consider the model in \cite{man2014uva,visentin2018uva} which is a system of  nonlinear differential equations (ODEs). In all equations, $t$ is the physical time (in min), all subscripts $b$ denotes basal state, and all of the parameters are given in the table \ref{tab:param}. The system of  nonlinear differential equations are given below:

{\it \textbf{Glucose Subsystem:}}
\begin{subequations}{\label{eq:ode1}}
	\begin{align}
	&\dot{G}_p(t)  = EGP(t)+Ra(t)-U_{ii} - E(t) - k_1 G_p(t) + k_2 G_t(t),&& G_p(0)  = G_{pb}\label{eq:ode1a}\\
	&\dot{G}_t(t)  = -U_{id} (t) + k_1G_p(t) - k_2 G_t(t), && G_t(0) = G_{tb}\\
	& G(t)  = \frac{G_p}{V_G}, &&  G(0) = G_b
	\end{align}
\end{subequations} 	
Here  $G_p$ (in mg/kg) is the mass of plasma glucose; $G_t$ (in mg/kg) is the mass of  tissue glucose; $G$ (in mg/dL) is  plasma glucose concentration and $V_g$ (in dL/kg) is the distribution volume of glucose; $EGP$ is the endogenous glucose production (in mg/kg/min); $Ra$ (in mg/kg/min) is the rate of glucose appearance in plasma; $U_{ii}$ (in mg/kg/min) and $U_{id}$ (in mg/kg/min) are insulin-independent and insulin-dependent glucose utilizations, respectively. Also $k_1$ and $k_2$ are the parameters.

{\it \textbf{Insulin Subsystem:}}
\begin{subequations}{\label{eq:ode2}}
	\begin{align}
	&\dot{I}_p(t)  = -(m_2+m_4)I_p(t) + m_1 I_l(t) + R_{ia}(t),&& I_p(0) = I_{pb}\\
	&\dot{I}_l(t)  = -(m_1+m_3)I_l(t) + m_2 I_p(t), && I_l(0) = I_{lb}\\
	&I(t)  = \frac{I_p(t)}{V_I}, && I(0) = I_b
	\end{align}
\end{subequations} 	
Here  $I_l$ (in pmol/kg) is the mass of liver insulin; $I_p$ (in pmol/kg) is the mass of  tissue insulin; $I$ (in pmol/L) is the plasma insulin concentration; $V_I$ (in L/kg) is the distribution volume of insulin; $R_{ia}$ (in pmol/kg/min ) is the rate of appearance of insulin in plasma; $m_1$, $m_2$, $m_3$ and $m_4$ are the parameters.

{\it \textbf{Glucose rate of appearance:}}
\begin{subequations}{\label{eq:ode3}}
	\begin{align}
	& Q_{sto}(t)  = Q_{sto1}(t) + Q_{sto2}(t),&& G_{sto}(0) = 0\\
	& \dot{Q}_{sto1}(t)  = -k_{gri}Q_{sto1}(t) + D \delta (t-\tau_D), && Q_{sto1}(0) = 0\\\label{eq:ode3_b}
	& \dot{Q}_{sto2}(t)  = -k_{empt}(Q_{sto})(t)Q_{sto2}(t) + k_{gri}Q_{sto1}(t), && Q_{sto2}(0) = 0\\
	& \dot{Q}_{gut}(t)  = -k_{abs}Q_{gut}(t) + k_{empt}Q_{sto}(t)Q_{sto2}(t), && Q_{gut}(0) = 0\\
	& Ra(t)  = \frac{f.k_{abs}.Q_{gut}(t)}{BW}, && Ra(0) = 0\\
	& k_{empt}(Q_{sto})  = k_{\min} + \frac{k_{\max}-k_{\min}}{2}. \\
	& \lbrace \tanh[\alpha(Q_{sto} - b.D)] - \tanh[\beta(Q_{sto} - c.D)] + 2 \rbrace
	\end{align}
\end{subequations} 	
Here $Q_{sto}$ (in mg) is the amount of glucose in the stomach, $Q_{sto1}$ (in mg) is the amount of liquid glucose in the stomach,  $Q_{sto2}$ (in mg) is the amount of solid glucose in the stomach, $Q_{gut}$ (in mg) is the glucose mass in the intestine; $D$ (in mg) is the amount of ingested glucose at time $\tau_D$; $BW$ (in kg) is body weight; $k_{empt}$ is the rate constant of the gastric emptying; $K_{gri}$, $k_{abs}$, $k_{max}$, $k_{min}$, $f$, $\alpha$, $\beta$ are the parameters.

{\it \textbf{Endogenous glucose production:}}
\begin{subequations}{\label{eq:ode4}}
	\begin{align}
	& EGP(t)  = k_{p1} - k_{p2} G_p(t) - k_{p3} X^L(t) + \xi X^H(t), && EGP(0) = EGP_b \\
	& \dot{I}'(t)  = -k_i \left[ I'(t) - I(t)\right] , && I'(0) = I_{b}\\
	& \dot{X}^L(t)  = -k_i \left[ X^L(t) - I'(t)\right],&& X^L(0) = I_b\\
	& \dot{X}^H(t) =  - k_H  X^H(t) +k_H \times \max \left[ H(t) - H_b,0\right], && X^H(0) = 0\label{eq:ode4_d}
	\end{align}
\end{subequations} 	
Here $X^L$ (in ) is delayed insulin action on $EGP$;  $X^H$ is delayed glucagon action on $EGP$; $I'$ is delayed insulin in compartment 1; $k_{p1}$, $k_{p2}$, $k_{p3}$, $\xi$, $k_i$, $k_H$ are the parameters.

{\it \textbf{Glucose utilization:}}
\begin{subequations}{\label{eq:ode5}}
	\begin{align}
	& U_{ii}(t)  = F_{cns}\\
	& U_{id}(t) = \frac{[V_{m0}+V_{mx}. X(t)] G_t(t)}{K_{m0} + G_t(t)}\\
	& \dot{X}(t) =  -  p_{2U} X(t) + p_{2U}[I(t) - I_b], && X(0) = 0
	\end{align}
\end{subequations} 
Here $U_{ii}$ (in mg/kg/min) and $U_{id}$ (in mg/kg/min) are insulin-independent and insulin-dependent glucose utilization; $X$ (in pmol/L) is insulin in interstitial fluid; $F_{cns}$, $V_{m0}$, $K_{m0}$, $p_{2U}$ are the parameters.

{\it \textbf{Renal excretion:}}
\begin{equation}{\label{eq:ode6}}
\begin{aligned}
& E(t)  = \begin{cases}
k_{e1} [G_p(t) - k_{e2}] & \text{if } G_p(t) > k_{e2}\\
0  & \text{if } G_p(t) \le k_{e2}
\end{cases}
\end{aligned}
\end{equation} 
Here $E(t)$ (in mg/kg/min) is the glucose renal exertion; $k_{e1}$ is the parameter.

{\it \textbf{Glucagon kinetics and secretion:}}
\begin{subequations}{\label{eq:ode8}}
	\begin{align}
	&\dot{H}(t) = -nH(t) + SR_H(t) +Ra_{H}(t), && H(0) = H_b \\
	&SR_H(t) = SR^s_H(t) + SR^d_H(t),\\
	&\dot{S}R^s_H(t)  = - \rho \left[ SR^s_H(t) - \max\left(\frac{\sigma[G_{th}-G(t)]}{\max(I(t) - I_{th},0)+1}+SR^b_H,0\right) \right],&& SR^s_H(0) = n H_b\\
	& SR_H^d(t)  = \delta \max \left( -\frac{dG(t)}{dt},0\right)
	\end{align}
\end{subequations} 	
Here $H$ (in ng/L) is the concentration of plasma glucagon; $SR_H$ (in ng/L/min) is the glucagon secretion;  $Ra_{H}$ (in ng/L/min) is the rate of appearance of glucagon  in plasma; $SR^s_H$ (in ng/L/min) and $SR^d_H$ (in ng/L/min) is the static and dynamic components of glucagon, respectively; $n$, $\rho$, $I_{th}$, $\delta$ are the parameters.

{\it \textbf{Subcutaneous insulin kinetics:}}
\begin{subequations}{\label{eq:ode7}}
	\begin{align}
	&R_{ia}(t) = k_{a1}I_{sc1}(t) + k_{a2}I_{sc2}(t) \\
	&\dot{I}_{sc1}(t)  = -(k_d+ k_{a1})I_{sc1}(t) + IIR(t),&& I_{sc1}(0) = I_{sc1ss}\\
	&\dot{I}_{sc2}(t)  = k_d.I_{sc1}(t)-  k_{a2}I_{sc2}(t),&& I_{sc2}(0) = I_{sc2ss}\\
	& IIR(t) = IIR_b + \frac{u_I(t)}{BW}
	\end{align}
\end{subequations} 	
Here $R_{ia}$ (in pmol/kg/min) is the rate of appearance of insulin in plasma; $I_{sc1}$ (in pmol/kg) is the amount of nonmonomeric insulin in the subcutaneous space; $I_{sc2}$ is the amount of monomeric insulin in the subcutaneous space;  $IIR(t)$ is the insulin infusion rate where $IIR_b$ is the basal infusion rate (in pmol/kg/min) from body and $u_I$ (in pmol/min) is the external insulin infusion rate; $k_{a1}$, $k_{a2}$, $k_d$ are the parameters. As the exogenous insulin infusion rate appears in the above equation in pmol/kg/min, we divide $u_I$ by the body weight $BW$ in the equation. Note that here the $u_I$ is in pmol/min. To convert the unit of insulin infusion rate $u_I$ from U/min to pmol/min, we multiply $u_I$ by 6944.4, that is the unit conversion is 1 U/min = 6944.4 pmol/min.

{\it \textbf{Subcutaneous glucagon kinetics:}}
\begin{subequations}{\label{eq:ode9}}
	\begin{align}
	&\dot{H}_{sc1}(t)  = -(k_{h1}+ k_{h2})H_{sc1}(t) + GIR(t),&& H_{sc1}(0) = H_{sc1ss}\\
	&\dot{H}_{sc2}(t)  = k_{h1}H_{sc1}(t)-  k_{h3}H_{sc2}(t),&& H_{sc2}(0) = H_{sc2ss}\\
	&Ra_{H}(t) = k_{h3}H_{sc2}(t) \\
	& GIR(t) = GIR_b + \frac{u_G(t)}{BV}
	\end{align}
\end{subequations} 	
\normalsize
Here $H_{sc1}$ (in ng/L) and $H_{sc2}$ (in ng/L) are the  glucagon concentration in the subcutaneous space; $IGR$ is the  glucagon infusion rate where $GIR_b$ is the basal glucagon infusion rate (in ng/L/min) from the body and $u_I$ is the external glucagon infusion rate (in ng/min);$k_{h1}$, $k_{h2}$, $k_{h3}$ are the parameters. As the exogenous glucagon infusion rate appears in the above equation in ng/L/min, we divide $u_G$ by the body volume $BV$ in the equation. Note that here the $u_G$ is in ng/min. To convert the unit of glucagon infusion rate from mg/min to ng/min, we multiply $u_G$ by $10^6$, that is the unit conversion is 1 mg/min = $10^{6}$ ng/min.

We write the  ODEs in Eqs.\ \eqref{eq:ode1}-\eqref{eq:ode9} in the form $\dot{\textbf{x}}(t) = \textbf{f}(\textbf{x}(t),\textbf{u}(t), \Theta_{Gb})$  where $\textbf{x} \in \mathbb{R}^{17}$ and $t$ is the physical time (in min). The variable $x_1$ represents  $G_p$, the mass of  glucose in plasma; the variable $x_2$ represents $G_t$, the mass of glucose in tissue; the variable $x_3$ represents the mass of liver insulin $I_l$; the variable $x_4$ represents the mass of plasma insulin $I_p$; the variable $x_5$ represents the amount of delayed insulin  $I'$ in compartment 1; the variable $x_6$ represents the amount of delayed insulin $X^L$ action on $EGP$; the variable $x_7$ represents the amount of solid glucose $Q_{sto1}$ in the stomach; the variable $x_8$ represents the amount of liquid glucose $Q_{sto2}$ in the stomach; the variable $x_9$ represents the glucose mass $Q_{gut}$ in the intestine; the variable $x_{10}$ represents the amount of interstitial fluid $X$; the variable $x_{11}$ represents the amount of static glucagon  $SR_H^s$;  the variable $x_{12}$ represents the amount of plasma glucagon  $H$; the variable $x_{13}$ represents the amount of delayed glucagon  $X^H$ action on $EGP$;  the variable $x_{14}$ represents the amount of nonmonomeric insulin  $I_{sc1}$; in the subcutaneous space; the variable $x_{15}$ represents the amount of monomeric insulin  $I_{sc2}$ in the subcutaneous space; the variable $x_{16}$ represents the amount of subcutaneous glucagon  $H_{sc1}$ in the subcutaneous space; the variable $x_{17}$ represents the amount of subcutaneous glucagon  $H_{sc2}$  in the subcutaneous space. 
Also $\textbf{u}(t) = [u_I(t) \quad u_G(t)]^T$, where $u_I$ is the external insulin and $u_G$ is the external glucagon. We define   $\Theta_{G_b}$ as the set of parameters for which the basal glucose level is $G_b$.

\subsection{Parameters}
\AS{There are a total of 46 parameters in  Eqs.\ \eqref{eq:ode1}-\eqref{eq:ode9}. The parameters are not given in \cite{man2014uva}. We set all the parameters  for `Glucose subsystem', `Insulin subsystem', `Glucose rate of appearance', `Endogenous glucose production', `Glucose utilization', `Glucose utilization', `Renal excretion', `Subcutaneous insulin kinetics' from the references  \cite{dalla2007meal,dalla2007gim}, except $k_{p1}, V_{m0}$ and $HE_b$. 
According to \cite{dalla2007gim}, the parameters are chosen to satisfy the steady-state  constraints in type I diabetes. 
The parameters $k_{p1}$ and $V_{m0}$ are set so that the steady state solutions provide the basal Glucose level $G_b$ and $EGP_b = 2.4$. In Type I diabetes, the endogenous glucose production is high \cite{dalla2007gim}, so we choose $EGP_b = 2.4$  mg/kg/min. We set  $IIR_b = 0$ and $GIR_b = 0$ as the model we consider is for Type I diabetes.
The commercial version of the UVA/Pavoda simulator \cite{T1DMSEG} allows  computing blood glucose responses to supplied dosages of insulin for some patients, but 
 does not provide all of the parameters.  We tune the parameter $HE_b$ so that the blood glucose response to insulin of the \textit{patient} we consider in this paper is qualitatively similar to the blood glucose response to insulin of a patient from the software  \cite{T1DMSEG} (adultaverage.mat). All of the parameters we use are listed in the Table \ref{tab:param} for reproducibility of the results. Our implementation of the model \cite{man2014uva}  has been published in GitHub \cite{DM2014}.}
 
 The equations for $k_{p1}$ and $V_{m0}$ are given below:

\begin{subequations}{\label{eq:param}}
	\begin{align}
	& k_{p1}  = EGP_b + k_{p2} G_{pb} + k_{p3}I_b\\
	& V_{m0} = \frac{(EGP_b - F_{cns}) (K_{m0}+G_{tb})}{G_{tb}}
	\end{align}
\end{subequations}

The basal steady states are given below: 
\begin{subequations}\label{eq:basal}
	\begin{align}
	& G_{pb} = G_b.V_g\\
	& G_{tb} = \frac{F_{cns}-EGP_b+k_1G_{pb}}{k_2} \\
	& I_{lb}  = I_{pb}.\frac{m_2}{m_1+m_3} \\
	& I_{pb} = \frac{IIR_{b}}{m_2+m_4-\frac{m_1m_2}{m_1+m_3}}\\
	& I_{sc1ss} = \frac{IIR_b}{k_d+k_{a1}}\\
	& I_{sc2ss} = \frac{k_d}{k_{a2}}I_{sc1ss}\\
	& SR^s_{Hb}= n H_b\\
	& H_{sc1ss} = \frac{GIR_b}{k_{h1}+k_{h2}}\\
	& H_{sc2ss} = \frac{k_{h1}}{k_{h3}}H_{sc1ss}
	\end{align}
\end{subequations}

Here, the basal values $G_b$ (in mg/dL), $IIR_b$ (in pmol/kg/min) and $GIR_b$ (in ng/L/min) are settable by the user.

\begin{table}[htbp]
	\begin{adjustwidth}{-2.25in}{0in}
		\centering
		\caption{Average parameters}
		{\setlength\doublerulesep{ 2pt}
			\begin{tabular}{lll}
				\toprule \midrule
				Parameter & Type I Value & Unit \\
				\midrule \midrule 
				$BW$  & 78 \cite{dalla2007meal} &  Kg\\
				$BV$  & 78 \cite{dalla2007meal} &  L\\
				$V_g$ &   1.49  \cite{dalla2007meal}	&	dL/kg \\
				$k_1$ &	 0.065 \cite{dalla2007meal,dalla2007gim} & $\min^{-1}$		\\
				$k_2$ &	 0.079 \cite{dalla2007meal,dalla2007gim} &	 $\min^{-1}$ 	\\
				$V_I$ &	 0.04 \cite{dalla2007meal}	 &	 L/kg\\
				$m_1$ &	 0.379  \cite{dalla2007meal}&	 $\min^{-1}$ 	\\
				$m_2$ &   0.673  \cite{dalla2007meal}& $\min^{-1}$		\\
				$m_4$ &	  0.269 \cite{dalla2007meal}&	$\min^{-1}$ 	\\
				$m_5$ &  0.0526 \cite{dalla2007meal} & min.kg/pmol		\\
				$m_6$ &	 0.8118 \cite{dalla2007meal}& dimensionless		\\
				$HE_b$ &   \AS{0.112}\cite{dalla2007meal}& dimensionless 	\\
				$k_{p1}$ &   change Eq.\ \eqref{eq:param}&	mg/kg/min  \\
				$k_{p2}$ &   0.0021 \cite{dalla2007meal,dalla2007gim}&	$\min^{-1}$ 	\\
				$k_{p3}$ &  0.009 \cite{dalla2007meal,dalla2007gim} &	mg/kg/min per pmol/L 	\\
				$k_{p4}$ &  0.0786 \cite{dalla2007meal} &	mg/kg/min per pmol/L  	\\
				$k_i$  &   0.0066 \cite{dalla2007meal}& $\min^{-1}$  	\\
				$k_{\max}$ 	& 0.0465 \cite{dalla2007meal} & $\min^{-1}$	\\
				$k_{\min}$	& 0.0076 \cite{dalla2007meal} & $\min^{-1}$ 	\\
				$k_{\text{abs}}$  & 0.023 \cite{dalla2007meal} & $\min^{-1}$		\\
				$k_{\text{gri}}$ 	&0.0465  \cite{dalla2007meal}&	$\min^{-1}$	\\
				$f$    &0.9  \cite{dalla2007meal} & dimensionless	\\
				$a$   &0.00016  \cite{dalla2007meal}& $\text{mg}^{-1}$		\\
				$b$   &0.68 \cite{dalla2007meal} & dimensionless	\\
				$c$  &0.00023 \cite{dalla2007meal} & $\text{mg}^{-1}$		\\
				$d$   &0.009 \cite{dalla2007meal} &	dimensionless  	\\
				$F_{cns}$ 	& 1  \cite{dalla2007meal}& mg/kg/min		\\
				$V_{m0}$  	&changes (Eq.\ \eqref{eq:param} & mg/kg/min	\\
				$V_{mx}$  	& 0.034 \cite{dalla2007meal} & mg/kg/min per pmol/L		\\
				$K_{m0}$  	&4661.21  & mg/kg	 \\	
				$P_{2u}$ 	&0.084 \cite{dalla2007meal} & $\min^{-1}$ 	\\
				$k_{e1}$ 	&0.0007  \cite{dalla2007meal} & $\min^{-1}$		\\
				$k_{e2}$ 	&269 \cite{dalla2007meal}& mg/kg		\\
				$k_d$   	&0.0164 \cite{dalla2007gim}& $\min^{-1}$	\\
				$k_{a1}$ 	&0.0018 \cite{dalla2007gim}&	$\min^{-1}$  	\\
				$k_{a2}$  	&0.0182 \cite{dalla2007gim} &	$\min^{-1}$ 	\\
				$\delta$ 	&0.682 \cite{man2014uva} & (ng/L per mg/dL)	 	\\
				$\sigma$ 	&1.093 \cite{man2014uva}& $\min^{-1}$		\\
				$n$ 	&0.15 \cite{man2014uva} & $\min^{-1}$		\\
				$\zeta$   	&0.009 \cite{man2014uva}&	(mg/kg/min per ng/L)	\\
				$\rho$  	&0.57 \cite{man2014uva}& (ng/L/min per mg/dL)		\\
				$k_H$  	&0.16 \cite{man2014uva}& $\min^{-1}$		\\
				$I_{th}$ 	&$I_b$ \cite{man2014uva} & (pmol/L)		\\
				$G_{th}$   	&$G_b$ \cite{man2014uva}& mg/dL	\\
				$k_{h1}$ 	&0.0164 \cite{man2014uva}& $\min^{-1}$		\\
				$k_{h2}$ 	&0.0018 \cite{man2014uva}& $\min^{-1}$	\\
				$k_{h3}$ 	& 0.0182 \cite{man2014uva}&	$\min^{-1}$ 	\\
				\midrule 

				\bottomrule
			\end{tabular}
		}
		\begin{flushleft} Average parameters.
		\end{flushleft}
		\label{tab:param}
	\end{adjustwidth}
\end{table}

\begin{table}[htbp]
	\begin{adjustwidth}{-2.25in}{0in}
		\centering
		\caption{Basal values}
		{\setlength\doublerulesep{2pt}   
			\begin{tabular}{lll}
				\toprule[1pt] \midrule
				Basal  & Type I Value& Unit \\
				\midrule \midrule 
				$X^H_b$   & 0 \cite{DMemail} & pmol/L	 	\\
				$EGP_b$  & 2.4 \cite{dalla2007gim} & mg/kg/min		\\
				$H_b$  & 93 \cite{DMemail} & ng/L	 	\\
			    $IIR_b$ &  0  \cite{dalla2007gim} & pmol/kg/min	 	\\
				\midrule \bottomrule[1pt]
				
			\end{tabular}
		}
		\begin{flushleft} Basal values.
		\end{flushleft}
		\label{tab:basal}
	\end{adjustwidth}
\end{table}

\section{Network Representation of the GIGM Model}

\AS{
In this section we construct  a \textit{graph} representation of the GIGM model. A graph
$\mathcal{G}(\mathcal{V}, \mathcal{E})$, consists of a set $\mathcal{V}=\{x_i\}$, $i = 1, \ldots, n$ of nodes and a set $\mathcal{E}$ of directed edges, where each edge is identified by an ordered pair $\{x_i, x_j\}$ of nodes $x_i, x_j \in \mathcal{V}$. }

\AS{
\textit{Complex networks} typically consist of two parts; a set of nodes with their interconnections which represent the topology of the network, and the dynamics which describes the time evolution of the network nodes. The prescribed dynamics of the nodes can be linear or non-linear.  We can construct a complex network from a dynamical system, simply by considering a state variable as a node and by drawing a directed edge from a node $x_i$ to another node $x_j$ if $x_i$ appears in the time derivative of $x_j$.
We call \textit{driver nodes}, the nodes which directly connect to the external inputs. We call \textit{target nodes},  the nodes  which  have prescribed state that must be satisfied at the final time \cite{klickstein2017energy}.
}
%

\begin{figure}[b!]
	\begin{adjustwidth}{- 1 in}{0in} 
		\includegraphics[width =1.15\textwidth]{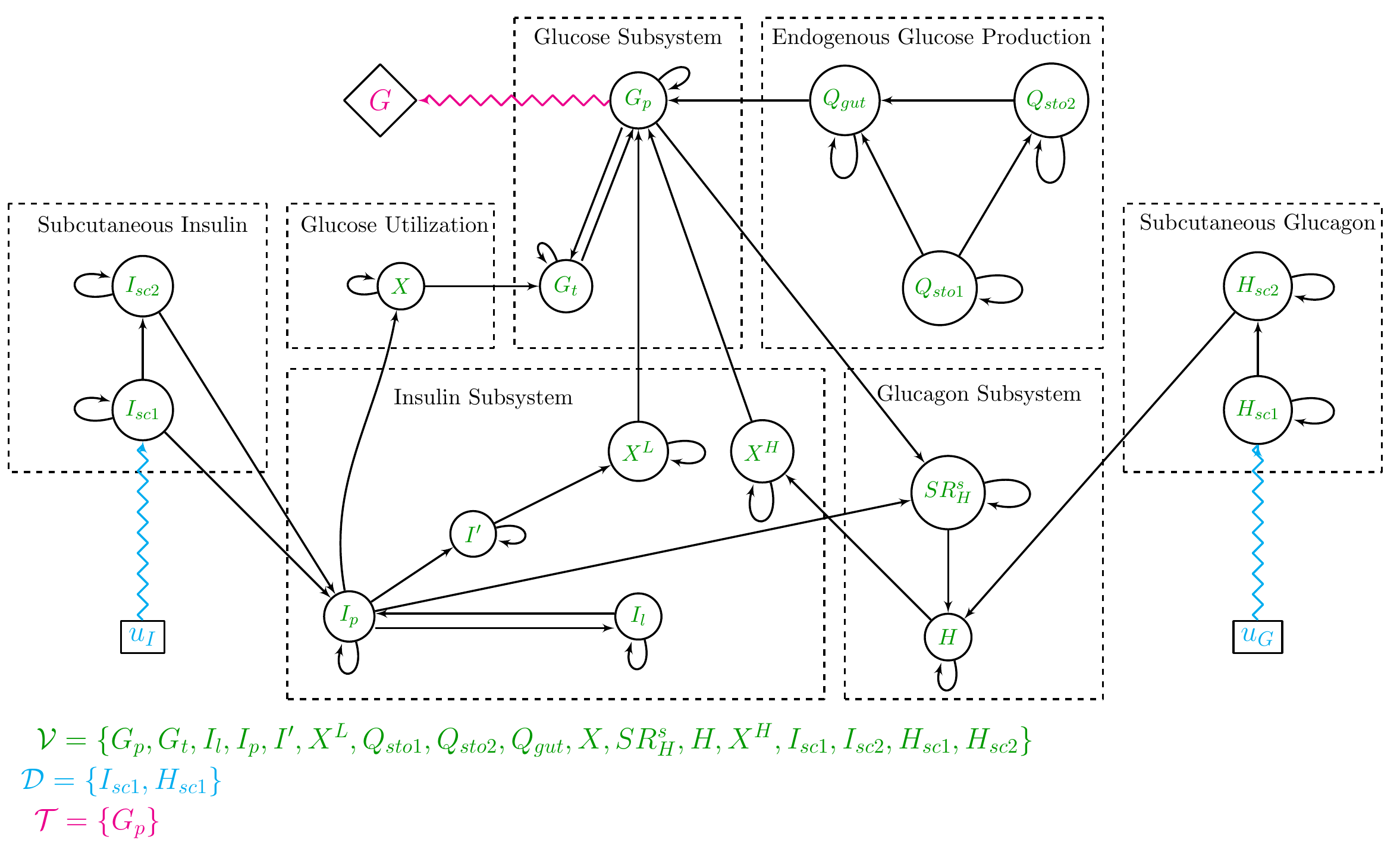}
		\caption{Network representation of the GIGM model.}
		\label{fig:network}
	\end{adjustwidth}
\end{figure}

\AS{ A network representation of the GIGM model is shown in Fig.\ \ref{fig:network}. 
Each one of the state variable $x_i$, $i = 1,\cdots,17$, in Eqs.\ \eqref{eq:ode1}-\eqref{eq:ode9}  is associated with a node (shown as a green circle in the figure). 
A directed edge (shown as a black arrow in the figure) is drawn from node $x_i$ to node $x_j$, if the state $x_i$ appears in the time derivative of the state $x_j$. For instance, as $G_t$ appears on the right hand side of Eq.\ \eqref{eq:ode1a}, there exists an edge from node $G_t$ to node $G_p$.  
%
In the model in Eqs.\ \eqref{eq:ode1}-\eqref{eq:ode9}, $u_I$ and $u_G$ are the external inputs acting on the nodes $I_{sc1}$ and $H_{sc1}$, respectively. Thus the set of drivers node $\mathcal{D}= \{I_{sc1},H_{sc1}\}$. In this model as the plasma glucose $G_p$ is the only variable we are trying to affect through the control action, thus the set of target nodes $\mathcal{T} = \{G_p\}$. In the figure, the driver nodes are colored cyan and the target node are colored magenta. As can be seen, the effect of the control inputs on the target node is mediated by the network structure, thus this particular structure plays an important role in our ability to control the network output. While the effect of the network topology has been investigated in the case of linear network dynamics, see e.g. \cite{liu2011controllability,ruths2014control,summers2014optimal,wang2012control,nepusz2012controlling,yuan2013exact,iudice2015structural,gao2016control,yan2015spectrum,yan2012controlling,sorrentino2007controllability,gates2016control}, the case of nonlinear networks has so far received only limited attention \cite{ cornelius2013realistic,wang2016geometrical,zanudo2017structure,klickstein2017locally}. Reference \cite{shirin2018prediction} and this paper investigate this problem in the context of two applications of interest to the medical field.
}

\section{Continuous Approximation of Non-differential Function in ODEs}\label{sec:approx}
\AS{The optimization algorithms implemented in $\mathcal{PSOPT}$ require the derivatives of the function $\textbf{f}(\textbf{x}(t),\textbf{u}(t), \Theta_{G_b})$ exists.} We notice that there are discontinuities in Eqs.\ \eqref{eq:ode1}-\eqref{eq:ode9}.  

\noindent The smooth approximation of the Renal exertion function $E(t)$ in Eq.\ \eqref{eq:ode6} by using a Heaviside function is,

\begin{equation}\label{eq:heavisidefunc}
E(t) = k_{e1}(G_p(t)-k_{e2})\times \mathcal{H}(G_p(t), k_{e2},k), 
\end{equation}
where,
\begin{equation}
\mathcal{H}(G_p(t), k_{e2},k) = \frac{1}{1+e^{-k(G_p-k_{e2})}}, k \in \mathbb{Z}.
\end{equation}
Here a larger $k$ corresponds to a sharper transition around $G_p(t) = k_{e2}$.

\noindent We define a continuous approximation of the Dirac delta function $\delta(t-\tau_D)$ in Eq.\ \eqref{eq:ode3_b},
\begin{equation}\label{eq:deltafunc}
\delta(t-\tau_D)= \frac{d}{dt}\mathcal{H}(t,\tau_D,k), 
\end{equation}
 where  $\mathcal{H}(t,\tau_D,k) = \frac{1}{1+e^{-k(t-\tau_D)}}, k \in \mathbb{Z}$. Here a larger $k$ corresponds to a sharper transition at $t = \tau_D$.

\noindent We also define continuous approximation of the $\max(.)$ function, e.g. in Eq.\ \eqref{eq:ode4_d}, as

\begin{equation}\label{eq:maxfunc}
\max(H(t) - H_b,0) = (H(t) - H_b)\times \mathcal{H}(H(t), H_b,k), 
\end{equation}
where  $\mathcal{H}(H(t), H_b,k) = \frac{1}{1+e^{-k(H-H_b)}}, k \in \mathbb{Z}$.
Here a larger $k$ corresponds to a sharper transition at $H(t) = H_b$. In all our approximation we set $k = 4$.

\end{document}